\documentclass[fleqn,usenatbib]{mnras}

\usepackage{newtxtext,newtxmath}

\usepackage[T1]{fontenc}

\usepackage{aecompl}

\usepackage{graphicx}	
\usepackage{amsmath}	
\usepackage{amssymb}	

\usepackage{xcolor}

\usepackage{subfig}
\usepackage{comment} 

\usepackage{etoolbox}
\makeatletter
\newcount\c@additionalboxlevel
\newcount\c@maxboxlevel
\patchcmd\@combinedblfloats{\box\@outputbox}{%
  \stepcounter{additionalboxlevel}%
  \box\@outputbox
}{}{\errmessage{\noexpand\@combinedblfloats could not be patched}}

\AtBeginShipout{%
  \ifnum\value{additionalboxlevel}>\value{maxboxlevel}%
    \typeout{Warning: maxboxlevel might be too small, increase to %
      \the\value{additionalboxlevel}%
    }%
  \fi 
  \@whilenum\value{additionalboxlevel}<\value{maxboxlevel}\do{%
    \typeout{* Additional boxing of page `\thepage'}%
    \setbox\AtBeginShipoutBox=\hbox{\copy\AtBeginShipoutBox}%
    \stepcounter{additionalboxlevel}%
  }%
  \setcounter{additionalboxlevel}{0}%
}
\makeatother

\newcommand{\quotes}[1]{``#1''}
\newcommand\code[1]{\textsc{\MakeLowercase{#1}}}

\def\be{\begin{equation}}
\def\ee{\end{equation}}

\def\cc{{\rm cm}^{-3}}
\def\kms{km s$^{-1}$}

\def\msolar{{\rm M}_{\odot}}
\def\msun{{\rm M}_{\odot}}
\def\zsun{{\rm Z}_{\odot}}
\def\lsun{{\rm L}_{\odot}}
\def\msunyr{\msun/{\rm yr}}
\def\msunpc2{\msun/{\rm pc}^{2}}

\def\tcmb{T_{\rm CMB}}

\def\lsim{\lesssim}

\def\CII{\hbox{[C~$\scriptstyle\rm II $]}}

\def\CIIion{\hbox{C~$\scriptstyle\rm II $}}

\def\kms{{\rm km\, s^{-1}}}

\def\althaea{Alth{\ae}a} 
\graphicspath{{plots/}} 

\title[Kinematics of early galaxies]{Kinematics of $z\geq 6$ galaxies from \CII~line emission}

\author[M. Kohandel et al.]{M. Kohandel$^{1}$\thanks{\href{mailto:mahsa.kohandel@sns.it}{mahsa.kohandel@sns.it}}, A. Pallottini$^{1,2}$, A. Ferrara$^{1}$,
A. Zanella$^{3}$, C. Behrens$^{4}$, S. Carniani$^{1}$,
\newauthor S. Gallerani$^{1}$, L. Vallini$^{5,6}$
\\
$^{1}$Scuola Normale Superiore, Piazza dei Cavalieri 7, I-56126 Pisa, Italy\\
$^{2}$Centro Fermi, Museo Storico della Fisica e Centro Studi e Ricerche \quotes{Enrico Fermi}, Piazza del Viminale 1, Roma, 00184, Italy\\
$^{3}$European Southern Observatory, Karl Schwarzschild Stra\ss e 2, 85748 Garching, Germany\\
$^{4}$Institut f\"{u}r Astrophysik, Georg-August Universit\"{a}t G\"{o}ttingen, Friedrich-Hundt-Platz 1, 37077, G\"{o}ttingen, Germany\\
$^{5}$Leiden Observatory, Leiden University, PO Box 9500, 2300 RA Leiden, The Netherlands\\
$^{6}$Nordita, KTH Royal Institute of Technology and Stockholm University Roslagstullsbacken 23, SE-106 91 Stockholm, Sweden\\
}

\date{Accepted XXX. Received YYY; in original form ZZZ}

\pubyear{2019}

\begin{document}
\label{firstpage}
\pagerange{\pageref{firstpage}--\pageref{lastpage}}
\maketitle

%
\begin{abstract}
We study the kinematical properties of galaxies in the Epoch of Reionization via the \CII 158$\mu$m line emission. The line profile provides information on the kinematics as well as structural properties such as the presence of a disk and satellites.
To understand how these properties are encoded in the line profile, first we develop analytical models from which we identify disk inclination and gas turbulent motions as the key parameters affecting the line profile.
To gain further insights, we use \quotes{\althaea}, a highly-resolved ($30\, \rm pc$) simulated prototypical Lyman Break Galaxy, in the redshift range $z = 6-7$, when the galaxy is in a very active assembling phase.
Based on morphology, we select three main dynamical stages: I)  Merger , II) Spiral Disk, and III) Disturbed Disk. We identify spectral signatures of merger events, spiral arms, and extra-planar flows in I), II), and III), respectively.
We derive a generalised dynamical mass vs. \CII-line FWHM relation. If precise information on the galaxy inclination is (not) available, the returned mass estimate is accurate within a factor $2$ ($4$).
A Tully-Fisher relation is found for the observed high-$z$ galaxies, i.e. $L_{\rm[CII]}\propto (FWHM)^{1.80\pm 0.35}$ for which we provide a simple, physically-based interpretation.
Finally, we perform mock ALMA simulations to check the detectability of \CII. When seen face-on, \althaea~is always detected at $> 5\sigma$; in the edge-on case it remains undetected because the larger intrinsic FWHM pushes the line peak flux below detection limit. This suggests that some of the reported non-detections might be due to inclination effects.
\end{abstract}



\section{Introduction}\label{sec:intro}

Answering the fundamental questions related to the formation, build-up, and mass assembly of galaxies is one of the main goals of modern astrophysics.
The first stars and galaxies formed when the diffuse baryonic gas in the Intergalactic Medium (IGM) was able to collapse into the potential well of the dark matter halos in the early universe.
The ultraviolet (UV) radiation produced by these first sources ionised the hydrogen atoms in the surrounding IGM. This process, called cosmic reionization \citep{Madau+1999,Gnedin+2000,Barkana+2001}, took about 1 billion years to reach completion at $z \sim 6$ \citep{Fan+06,McGreer+11}.
After the formation of first sources, as time progressed, those objects gradually evolved, merging with their neighbours and accreting large quantities of gaseous fuel from a filamentary IGM. Then, through a combination of galaxy-galaxy mergers, rapid star formation, and secular evolution, the morphology of those galaxies transformed into what is observed locally. Both observationally and theoretically, understanding the details of the assembly process has proven very challenging as the internal structure of these system should be resolved.

Integral field spectroscopy and adaptive optics technology have enabled us to obtain diagnostic spectra of spatial regions resolved on scales of roughly $\sim 1 \, \rm kpc$ at intermediate redshifts ($z\sim 2-3$). These remarkable experiments revealed that such galaxies have  irregular and clumpy morphologies while their velocity structures are often consistent with rotating disks \citep{Genzel+11,Forster-schreiber+18,leung:2019}. The question remains if the situation is the same for the galaxies at even higher redshifts.

Over the last few years, observations have also managed to probe galaxies at progressively higher redshifts (for a recent review see \citealt{Dayal18}), producing a first, albeit partial, census of galaxy populations well into the Epoch of Reionization (EoR).
Although with UV surveys \citep[e.g.][]{Smit+14,Bouwens+15} the discovery of such galaxies has become possible, physical insights on the properties of the Interstellar Medium (ISM) of these sources rely on the detection of far-infrared (FIR) lines. It has now become possible with the advent of the Atacama Large Millimeter Array (ALMA) to detect these emission lines from high-$z$ galaxies.

Among the FIR lines, the fine-structure transition $2P_{3/2} \rightarrow 2P_{1/2}$ of singly ionised carbon at $\lambda = 158 \mu \rm m$ is the brightest one, accounting for $0.1 \%$ to $1\%$ of the total FIR luminosity \citep{Stacey+1991}, making it as one of the most efficient coolants of the ISM \citep{Malhotra+1997,Luhman+1998,Luhman+03}.
Neutral carbon has a relatively low ionisation potential ($11.3\, \rm{eV}$) and its distinctive line transition (\CII) is very easy to excite ($E/k \approx 92\, \rm K$). These properties are such that the line can arise from nearly every phase in the ISM. It can emerge from diffuse HI clouds, diffuse ionised gas, molecular gas and from the photodissociation regions (PDRs).
So far, the \CII$158 \, \mu \rm m$ line has been measured in a rapidly increasing number of galaxies at $z>6$ \citep[e.g.][]{Maiolino+15,Capak+15,Pentericci+16,Carniani+17,Jones+17,Matthee+17,Smit+18,Carniani+18clumps,Carniani+18himiko}.

Alongside observations, theoretical attempts have been made to model the \CII~emission and interpret the observations at $z>6$ \citep{vallini:2013,vallini:2015,pallottini:2017dahlia,Olsen+17,Katz+19} using numerical simulations of galaxies. So far, the purpose of theoretical modellings was mostly to estimate the total \CII~luminosity of galaxies at the EoR and understanding the relative contribution from different ISM phases. These theoretical works agree on the fact that most of the total \CII~luminosity arises from the dense PDRs \citep[][]{pallottini:2017dahlia} with a slight dependence on galaxy mass \citep{Olsen+17}.
Still no clear consensus has been reached whether or not the local \CII~star formation rate (SFR) relation that is observed locally \citep{DeLooze+14} holds for $z>6$ galaxies \citep[cfr][]{Carniani+18clumps}.
For instance while \citet{vallini:2015} and \citet{pallottini:2017dahlia} show that a deviation is present, \citet{Katz+19} show that for their suite of simulations at $z\sim 9$, the local relation holds.
The \CII~-SFR relation is further analysed in different works \citep{ferrara:2019,pallottini:2019}, where it is connected to galaxy evolutionary properties.

With the improvement of the quality of the view that ALMA is giving us from the high-$z$ universe, the \CII~line is starting to be considered as a suitable tool for studying the gas kinematics as well.
For instance, \citet{Smit+18} recently presented \CII~observations of two galaxies at $z\sim 7$ characterised by velocity gradients consistent with undisturbed rotating gas disks. Also in \citet{Jones+17}, using the \CII~line emission from a $z\sim 6$ Lyman break galaxy, conjectured that their observed system represents the early formation of a galaxy through the accretion of smaller satellite galaxies along a filamentary structure.
However, the build-up process, kinematics, and morphology of these galaxies are almost uncharted territories. Also, whether a disk structure is expected at those early epochs and whether it can survive the frequent collisions with merging satellites and accreting streams are key questions for galaxy formation theories.

In this work, we explore these questions by modelling the spectral profile of the \CII~emission coming from galaxies at $z>6$. To this aim, we first construct a simple galaxy model with controllable parameters and study the emerging \CII~spectra. Then, we trace the evolution of a prototypical Lyman Break Galaxy (LBG) -- \quotes{Alth{\ae}a} \citep{pallottini:2017althaea} -- from $z=7$ to $z=6$ through its \CII~emission maps and corresponding synthetic spectral line profiles.

The paper is organised as follows. In Sec. \ref{sec02}, we detail the emission model used throughout the paper, in particular analysing the effects of various assumptions made; this is followed by the description of our analytical galaxy model (Sec. \ref{sec03}) and the corresponding results. Then in Sec. \ref{sec04}, the description of the hydrodynamical simulation used in this work is given, along with the results obtained by combining it with our emission model. Then in Sec. \ref{sec05}, we compare our findings with the available \CII~observations. Finally, conclusions are summarised in Sec. \ref{sec06}.

\section{\CII~emission model}\label{sec02}

The \CII~transition can be excited via collisions of singly ionised carbon atoms (\CIIion) with other species present in the gas. Following \citet{dalgarno:1972}, we consider a partially ionised volume of gas in which carbon atoms are maintained in \CIIion~stage by far UV radiation in the Habing band ($6< {\rm h}\nu/{\rm eV} < 13.6$, \citealt{habing:1968}). The \CII~emissivity ($\varepsilon$), excited by collisions with free electrons and hydrogen atoms, is written as a function of the gas ($n$), electron  ($n_e$), and neutral hydrogen ($n_H$) number densities as follows:
\begin{equation}\label{eqn:ciiemissivity}
\varepsilon(n, T) =n\, (\frac{Z}{\zsun})\, A_{C}\left[\frac{n_H}{1+{n_H}/{n_H^{\rm cr}}}\Lambda^{H} + \frac{n_e}{1+{n_e}/{n_e^{\rm cr}}}\Lambda^{e}\right]\, ,
\end{equation}
where $\Lambda^{H}=\Lambda^{H}(T)$ and $\Lambda^{e}=\Lambda^{e}(T)$ are the specific cooling rates due to collision with $H$ atoms and free electrons at temperature $T$. $Z$ is the metallicity of the gas, $\zsun =0.0134$ is the solar metallicity \citep{asplund:2009}, and $A_C = 2.69 \times 10^{-4}$ is the adopted solar ratio of carbon to hydrogen number densities \citep{asplund:2009}.
Note that we have included in an approximate manner the effects of the critical density $n_H^{cr} = 3000\, \cc$ and  $n_e^{cr} = 8\,\cc$ \citep{Goldsmith+12} for hydrogen and electron collisions to ensure the validity of eq. \ref{eqn:ciiemissivity} in high density regimes\footnote{For each type of collision partner, the critical density $n_{cr}$ is defined by the collisional de-excitation rate being equal to the effective spontaneous decay rate. If the density is well below $n_{cr}$, one can use the \citet{dalgarno:1972} definition for cooling rate as the product of singly ionised carbon density with hydrogen/electron number density.}.

We require $\varepsilon$ to vanish in highly ionised regions ($T>10^4 \, \rm K $) where our assumption that all the carbon is singly ionised would not be valid anymore.
In this treatment we also assume that the \CII~line is optically thin \citep[see discussion in][]{Goldsmith+12}, which means that the integrated intensity is proportional to the \CIIion~ column density along the line of sight (l.o.s.), irrespective of the optical depth of the medium (see also Sec. \ref{sec_sim_couple}). In this approximation, for each gas parcel of volume $V$, we then compute the \CII~luminosity as $L=\varepsilon V$.

\subsection{CMB effects}
\begin{figure}
\centering
\includegraphics[width=0.5\textwidth]{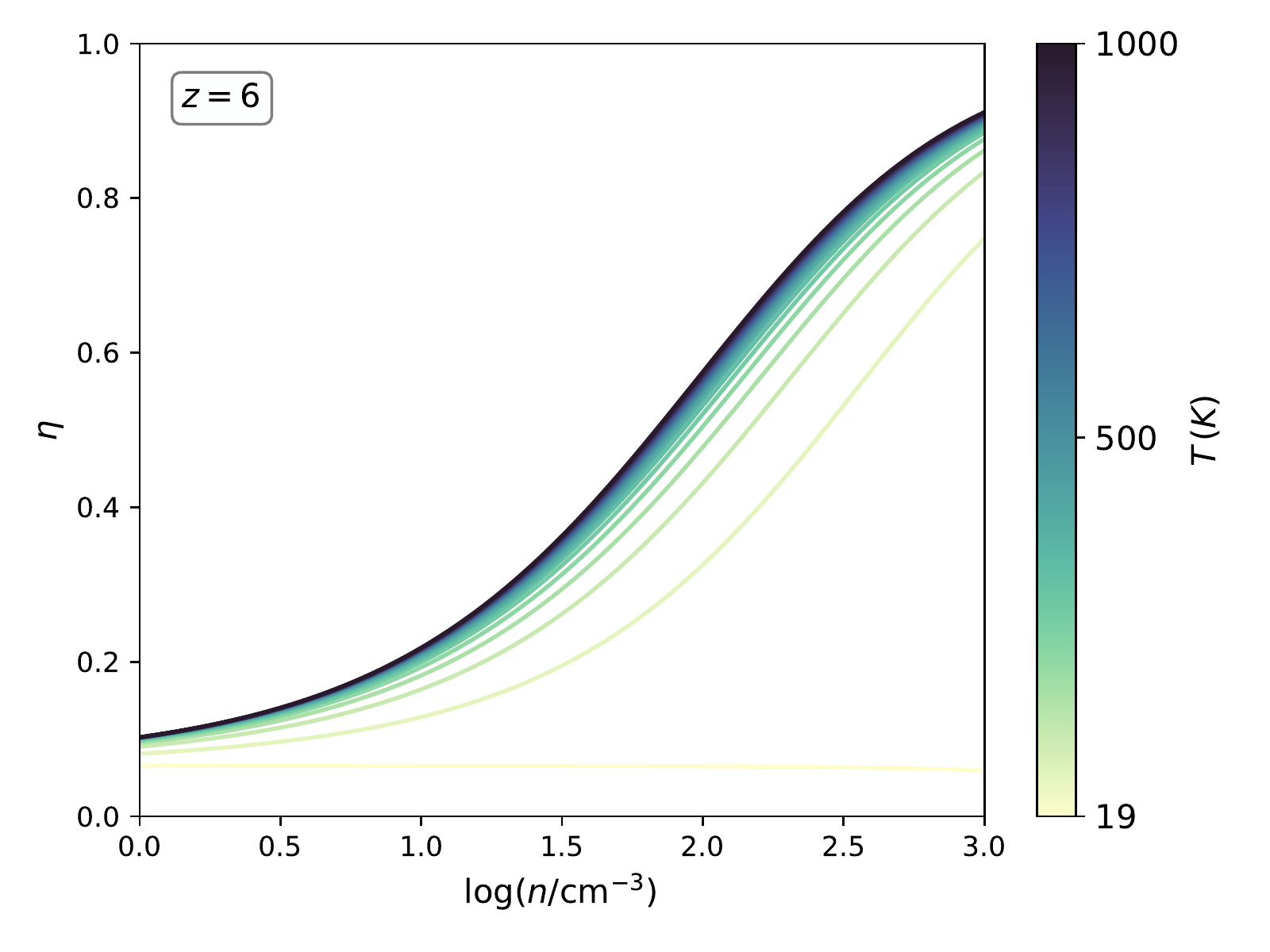}
\caption{
CMB suppression of \CII~emission ($\eta$) as a function of gas number density ($n$). Different lines indicate different gas temperature ($T$). The suppression is calculated at redshift $z=6$ via eq. \ref{eq:suppression_base_def}; see the text for the details of the calculation. \label{fig01}
}
\end{figure}

The Cosmic Microwave Background (CMB) has a thermal black body spectrum at a local temperature of $\tcmb^{0} = 2.725$ K, increasing with redshift as $\tcmb(z) = (1+z)\,\tcmb^{0}$. Assuming local thermal equilibrium, this sets the minimum temperature of the ISM, which at high redshift becomes non-negligible. Any emission coming from the ISM will be seen against the CMB background. As discussed in \citet{DaCunha2013}, the contrast of the emission against the CMB radiation in the rest-frame is given by:
\begin{equation}
\Delta I_{\nu} = [B_{\nu}(T_s)-B_{\nu}(\tcmb)]\left(1-e^{-\tau}\right)\,,
\end{equation}
where $B(\nu)$ is the Planck function and $T_s$ is the spin temperature of the FIR line.
Assuming the \CII~ line to be optically thin in the sub-mm band, i.e. $e^{-\tau_\nu} \approx 1-\tau_{\nu}$, the ratio between the flux observed against the CMB and the intrinsic flux emitted will be\footnote{FIR flux observed against CMB is defined as $F^{\rm{obs}}_{\nu/(1+z)}=(1+z) A\Delta{I_{\nu}}/d_{L}^2$, where $A$ is the physical area of the galaxy and $d_{L}$ is the luminosity distance.}:
\begin{equation}\label{eq:suppression_base_def}
\eta \equiv \frac{F^{\rm{obs}}_{\nu/(1+z)}}{F^{\rm{int}}_{\nu/(1+z)}}= 1-\frac{B_{\nu}(\tcmb)}{B_{\nu}(T_s)}\,.
\end{equation}

As $T_s$ approaches $\tcmb$, $\eta \rightarrow 0$; in this case the CMB completely suppresses the line flux. For \CII, the spin temperature is defined using the ratio of the thermal equilibrium population of the upper  ($u$: $2_{P_{3/2}})$, and lower ($l$: $2_{P_{1/2}}$) level of fine structure transition:
\begin{equation}\label{eq:poplevel}
\frac{n_{u}}{n_{l}}=\frac{g_{u}}{g_{l}}e^{-T_*/T_s} \, ,
\end{equation}
where $T_* = 91.7 \, \rm{K} $ is the equivalent temperature of the level transition, and $g_u = 4$, $g_l = 2$ are the statistical weights.
Following the procedure used in \citet{vallini:2015} \citep[see also][]{Pallottini+15}, $T_s$ is defined as:
\begin{equation}\label{eq:spin_temperature}
\frac{T_*}{T_s} = \ln{\frac{A_{ul}(1+\frac{c^2 I_\nu}{2h\nu})+n_eC_{ul}^{e}+n_HC_{ul}^{H}}{A_{ul}(\frac{c^2 I_{\nu}}{2h\nu^3})+n_eC_{ul}^ee^{-T_*/T}+n_HC_{ul}^{H}e^{-T_*/T}}}\,,
\end{equation}
where $A_{ul}$ is the Einstein coefficient for spontaneous emission and $C_{ul}^{e}$ ($C_{ul}^H$) is the collisional de-excitation rate for collisions with $e$ (H-atoms). For the \CII~ line emission $A_{ul} = 2.36 \times 10^{-6} \, \rm s^{-1}$ \citep{Suginohara+1999} and $C_{lu}^{e}(T)=(8.63 \times 10^{-6}/{g_l \sqrt{T}})\gamma_{lu}(T)e^{-T_{*}/T}$ with $\gamma_{lu}(T)$ being the effective collision strength computed based on \citet{Keenan+1986}. $C_{lu}^{H}(T)$ is tabulated in \citet{dalgarno:1972}.

As discussed in \citet{gong:2012}, at high redshifts the soft UV background at $1330$ \AA~produced by the first galaxies and quasars can pump the \CIIion~ ions from the energy level $2s^22p \,\, 2P_{1/2}$ to $2s2p^2 \,\, 2D_{3/2}$ ($\lambda=1334.53$ \AA~), and $2s^22p \,\, 2P_{3/2}$ to $2 s^2 2p^2\,\, 2D_{3/2}$  ($\lambda=1335.66$ \AA). This pumping effect can lead to the \CII~transition $2D_{3/2} \rightarrow 2P_{3/2} \rightarrow 2 P_{1/2}$ which would mix the levels of the \CII~line. Similarly to \citet{vallini:2015}, we add this UV pumping effect in eq. \ref{eq:spin_temperature}.

To summarise, with $n_e$, $n_H$ and $T$ we can compute the spin temperature of \CII~ line using eq. (\ref{eq:spin_temperature}) and the CMB suppression using eq. (\ref{eq:suppression_base_def}). In Fig. \ref{fig01}, the CMB suppression factor, $\eta$, is shown as a function of gas density for different temperatures and for $z=6$. We fix the metallicity to be $Z=0.5 \, \zsun$ and vary the temperature. \footnote{ In this case the ionisation fraction of the gas is computed by solving the equilibrium between collisional ionisation, ionisation due to cosmic rays and X-rays and recombination rates for H and He \citep{wolfire+95}. It depends on hydrogen number density, temperature, and metallicity of the gas.} The cooler the gas, the more the \CII~emission is suppressed. Note that, independently of $T$, the emission is suppressed by about 90\% for low-density gas ($n\lsim 1\,\cc$), because collisions are not efficient enough to decouple $T_s$ from the temperature of the CMB, in agreement with results in the literature \citep{gong:2012,vallini:2015,Pallottini+15}.

\section{Semi-Analytical insights}\label{sec03}

\begin{figure*}
\centering
\subfloat{\includegraphics[width=0.48\linewidth]{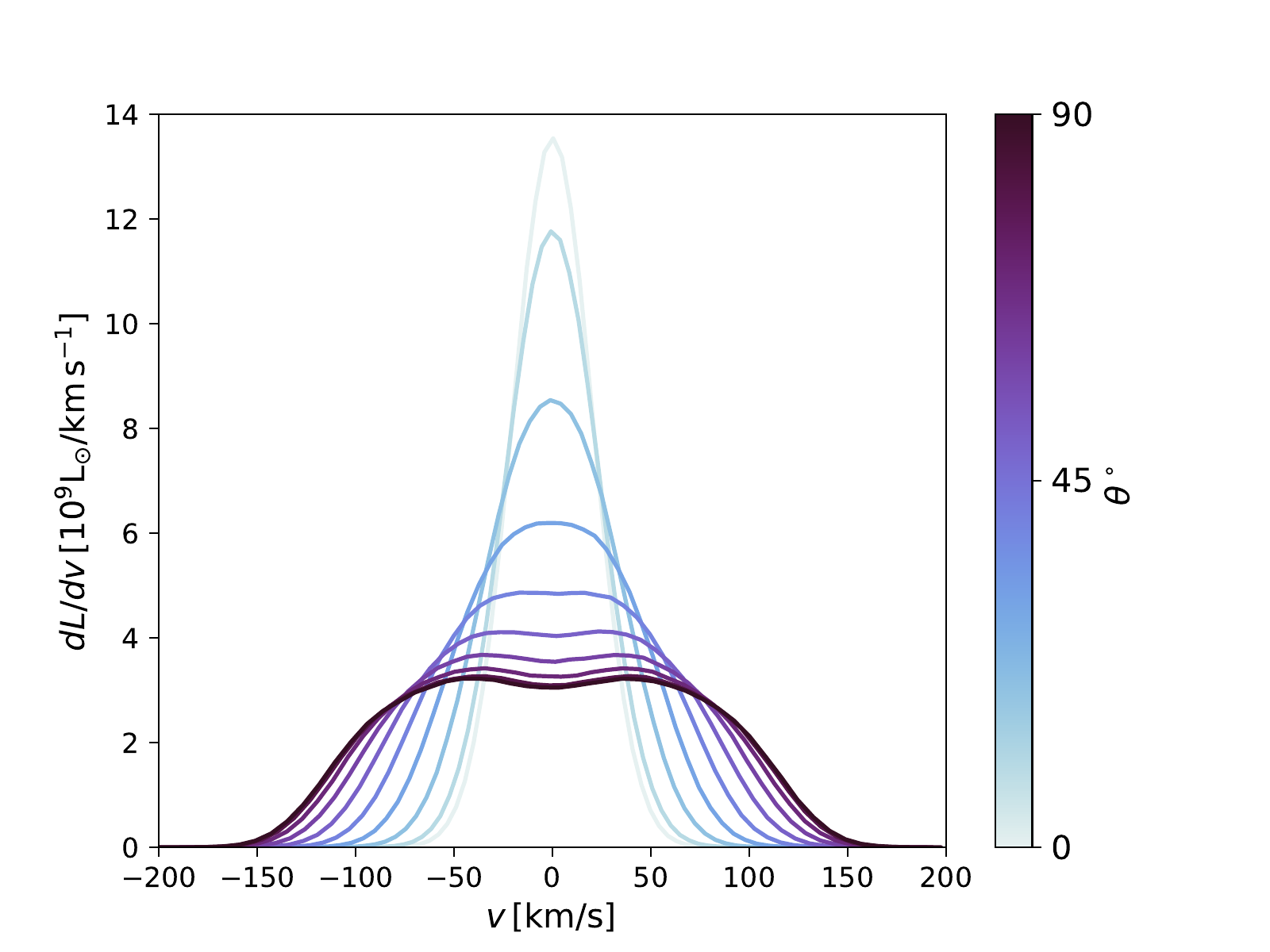}}
\hfill
\subfloat{\includegraphics[width=0.48\linewidth]{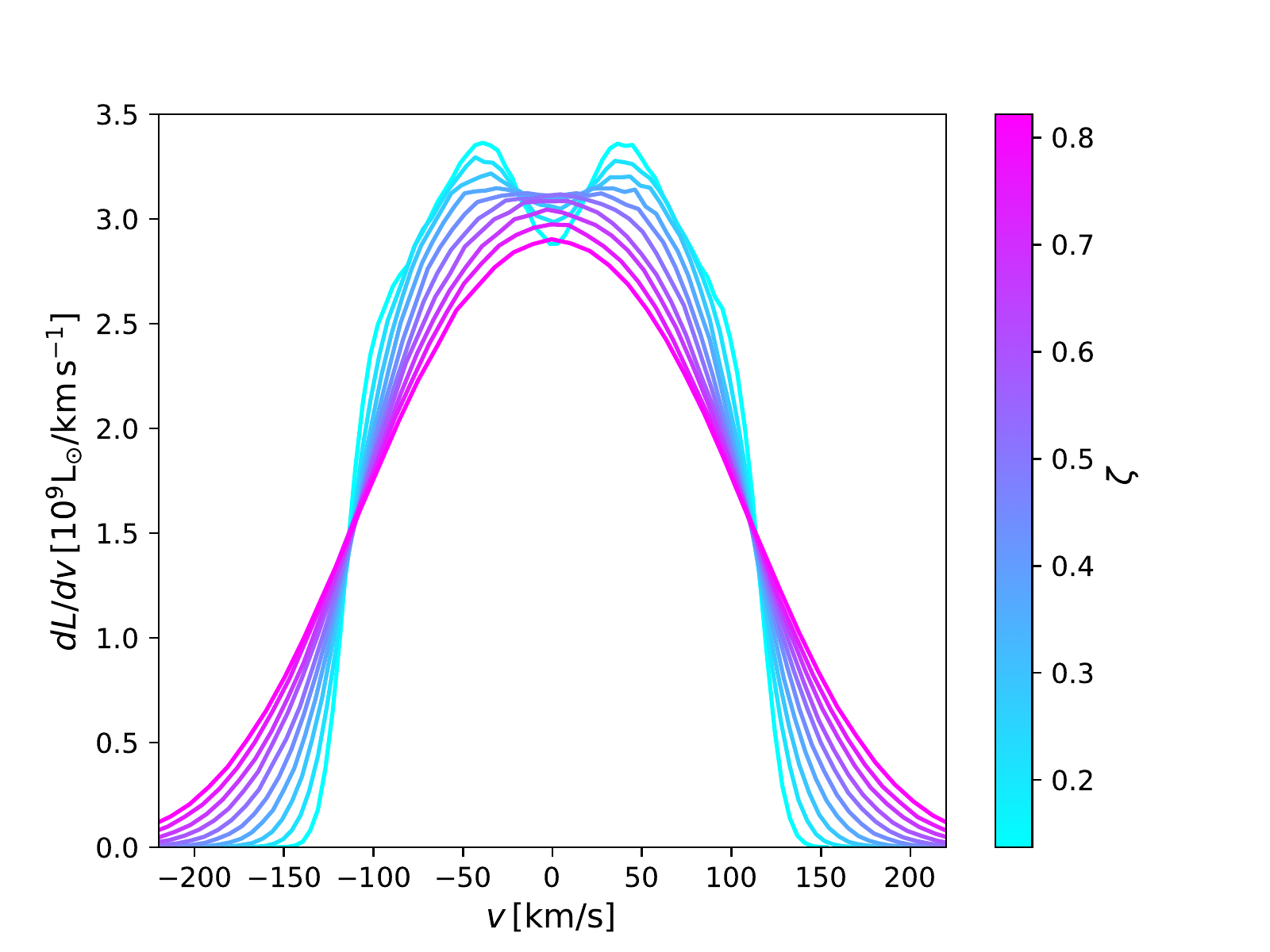}}
\caption{ Spectral profile of a geometrically thin disk with an exponential profile. {\it Left panel}: Disk inclination is between $\theta=0^{\circ}$(face-on) and $\theta=90^{\circ}$ (edge-on). In addition to rotational velocities, isotropic turbulent motions extracted from a Gaussian distribution having an r.m.s. amplitude of $20\,\kms$ are added to the disk. {\it Right panel}: The inclination of the disk is fixed to be edge-on and we vary the amplitude of turbulent motions. Both inclination and large turbulent motions broaden the wings of the line and lower the amplitude by a factor of $\sim 4 $ and $\sim 1.2$  respectively.
\label{fig03}}
\end{figure*}

We start by developing a simple analytical model of a disk galaxy to elucidate the physics involved in shaping the line profile, and to build a controlled environment for the analysis of \CII~emission from high-$z$ galaxies.
We consider a geometrically-thin disk and assume that the surface-brightness profile of the disk has an exponential form:
\begin{equation}
I(r) \propto \exp(-r/r_d) \, ,
\end{equation}
where $r_d$ is the disk scale length. If the mass surface  density is also exponential with the same scale length, i.e.:
\be\label{eq_surfac_density}
\Sigma(r)=\Sigma_0 \exp{(-r/r_d)}\,,
\ee
the potential that such a disk would generate at the equatorial plane is \citep{binney:2008book}:
\begin{equation}\Phi(r,0)=-\pi G \Sigma_0 r [I_0(y)K_1(y)-I_1(y)K_0(y)] \, ,
\end{equation}
where $y={r}/{2r_d}$ and $I_n$, $K_n$ are the modified Bessel functions of first and second kind, respectively. If we differentiate this potential with respect to $r$, we obtain the circular speed of the exponential disk \citep{Freeman+1970}:
\begin{equation}\label{exponentialvel}
v_c^2(r)=4\pi G \Sigma_0 r_d y^2 [I_0(y)K_0(y)-I_1(y)K_1(y)]\, .
\end{equation}
Using the circular velocity $v_c$, we can define the velocity along the l.o.s. as follows:
\begin{equation}\label{eq_exponentail_vel_second}
v(r,\theta, \phi)^2 = 4\pi G \Sigma_0 r_d y^2 [I_0(y)K_0(y)-I_1(y)K_1(y)]\cos^2{\phi} \sin^2{\theta} \,,
\end{equation}
where $\theta$ is the angle between the l.o.s axis and the normal to the disk plane and $\phi$ is the polar angle on the plane of the face-on disk. We assume a thin disk with $\Sigma_0 = 1000\,\msunpc2$, $r_d=3\, {\rm kpc}$, and a thickness of $100\, {\rm pc}$.

For our kinematic analysis, it is useful to define a 2D Cartesian grid centred on the galaxy centre. We choose a grid of size $(24 \, \rm kpc)^2$ divided in a total of $(4\times10^3)^2$ cells, i.e. each cell has a linear resolution of $6\, \rm pc$. In each cell, surface density and velocities are computed using eqs. \ref{eq_surfac_density} and \ref{eq_exponentail_vel_second}, respectively. We also account for random turbulent motions (i.e. deviations from perfect circular orbits) by adding in each cell a random velocity, the components ($t$) of which are extracted from a Gaussian distribution:
\begin{equation}
p(t)=\frac{1}{\sqrt{2\pi {v_t}^2}}e^{-t^2/2{v_t^2}}\,,
\end{equation}
where $v_t$ is the standard deviation of the distribution. We further assume isotropic turbulence so the three added components have the same magnitude.

Assuming a uniform temperature of $100 \, \rm K$\footnote{The reference temperature $T=100\,K $ is the mean temperature found for molecular gas in our high-$z$ galaxies simulations, see Fig. 8 in \citet{pallottini:2017althaea}.} for the disk, an ionisation fraction of $x_e=0.2$, and a metallicity of $Z=\zsun$, we compute the \CII~luminosity using the model described in Sec. \ref{sec02}. Having the l.o.s velocity and luminosity for each cell, we extract the integrated spectral profile by computing the histogram of velocities weighted by the corresponding value of \CII~luminosity.

First, we explore the effect of inclination of the disk by focusing on the spectral profile of the emission. In the left panel of Fig. \ref{fig03}, we show the \CII~spectra from our disk galaxies including turbulent velocities with $v_t=20 \,\kms$ . Different lines correspond to a different inclination of the disk. As discussed by \citet{Elitzur+12}, the spectral profile of such a disk in the edge-on view ($\theta = 90^\circ$) should show a double peak structure. We see in Fig. \ref{fig03} that inclining the disk from face-on view ($\theta = 0$) to the edge-on one smoothly changes the spectral profile from having a Gaussian shape to the double peak structure. Also inclining the disk towards edge-on produces broader wings compared to the face-on case. In addition, the peak amplitude of the line decreases by factor of $\sim 4$ in the edge-on case. These effects happen because by inclining the disk towards the edge-on view, $\sin(\theta)\rightarrow 1$ (see eq. \ref{eq_exponentail_vel_second}) allows for stronger contributions from high l.o.s. velocities. Consequently, the peak amplitude of the line decreases to keep the total \CII~luminosity, given by the integral below the curve, constant.

Random motions also change the spectral profile. In the right panel of Fig. \ref{fig03}, we set the inclination of the disk to be edge-on (double peak profile) and then vary $v_t$. For each of the cases with different turbulence velocities, we calculate $\zeta = v_t/\bar{v}_c$ in which $\bar{v}_c \simeq 75\,\kms$ is the mass-weighted average circular velocity of the exponential disk. We find  (Fig. \ref{fig03}, right panel) that if $\zeta > 0.5$ the double peak profile is erased, which means that turbulent motions can mask the presence of the disk in the spectrum. Furthermore, and similarly to the effect of inclination discussed above, turbulence broadens the line wings and decreases the line intensity at the peak by a factor of $\sim 1.2$.

With these controlled case examples, we conclude that depending on the inclination of the disk and the amount of turbulent motions, emission from a rotating disk might produce quite a range of different line profiles. In particular, inclination and turbulence have a degenerate effect in changing the spectral shape of emission. The double peak signature of our rotating edge-on disk is erased either by changing the inclination ($\theta<70^{\circ}$) or significant turbulent velocities ($\zeta > 0.5$). Similarly, the single Gaussian shape can be the signature of a highly turbulent disk or simply a face-on view of a disk with moderate turbulent motions.

Here, for a better comparison with the following analysis of the simulation (Sec. \ref{sec04}), it is convenient to define two cases of our analytical model; Smooth Disk: a smooth disk with $\zeta < 0.5$ featuring a symmetric double-peak profile in the edge-on view and a single Gaussian profile in the face-on view and Turbulent dominated Disk: a Disturbed Disk with $\zeta > 0.5$, which has a smooth single Gaussian spectral profile both in the face-on and edge-on view.

\section{High redshift galaxy simulations}\label{sec04}

We now turn our analysis to the more realistic case of galaxies extracted from zoom-in cosmological simulations, whose main features are outlined below. This is a necessary step to produce reliable predictions that catch the ISM complexity during galaxy assembly and thus can be directly confronted with observational data.
For details of the simulation, we refer the reader to \citet[][]{pallottini:2017althaea}.

\citet{pallottini:2017althaea} uses a customised version of the adaptive mesh refinement code \code{ramses} \citep{Teyssier+02} to zoom-in on the evolution of \quotes{Alth{\ae}a}, a $z\sim 6$ LBG hosted by a dark matter halo of mass $\simeq 10^{11} \msolar$. The gas mass resolution of the zoom-in region in this simulation is $1.2 \times 10^4 \msolar$ and the additional adaptive refinement allows us to resolve spatial scales down to $\simeq 30 \, \rm{pc}$ at $z \sim 6$.
In this simulation, a non-equilibrium chemical network has been implemented via the code \code{KROME} \citep{Grassi+14} which includes $\rm{H}$, $\rm{H}^{+}$, $\rm{H}^{-}$, $\rm{He}$, $\rm{He}^{+}$, $\rm{He}^{++}$, $\rm{H}_2$, $\rm{H}_2^{+}$, and electrons \citep[see also][]{Bovino+16}.
Stars are formed according to the Kennicutt-Schmidt relation \citep{Schmidt+95,Kennicutt+98} that depends on the molecular hydrogen density computed from the non-equilibrium chemical network.
As described in \citet{pallottini:2017dahlia}, stellar feedback includes supernovae, winds from massive stars, and radiation pressure. It also accounts for the blast wave evolution inside molecular clouds. The thermal and turbulent energy content of the gas is modelled similarly to \citet{Agertz+15}.

At $z=6$, \althaea~is characterised by a stellar mass $M_{\star}\sim 10^{10}\msun$ and $\rm{SFR}\sim 100\, \msunyr$. During its evolution, the SFR-stellar mass relation of \althaea~is comparable to what is inferred from high-$z$ observations \citep[][]{jiang:2016}.
By modelling the internal structure of molecular clouds, \citet{vallini:2018} used \althaea~to predict the CO line emission.
By post-processing the simulation with radiation transfer through dust \citet{behrens:2018dust} were able to reproduce the observed properties of A2744\_YD4 \citep{laporte:2017}, one of the most distant ($z \approx 8.3$) galaxies where dust continuum is detected.

In this work we are interested in studying the evolution of \althaea~from $z=7$ to $z=6$ in its integrated \CII~surface brightness (luminosity), and the corresponding spectra along different l.o.s. identified by $\hat{n}$. Our aim is to investigate different kinematical features and their connection with the assembly process as imprinted in the \CII~line profile.

\subsection{Computing \CII~maps and spectra}\label{sec_sim_couple}

The first step is to compute the \CII~luminosity. For that, we need $n$, $n_H$, $n_e$,  $T$ and $Z$ as the inputs for the emission model (eq. \ref{eqn:ciiemissivity}) and CMB suppression (eq. \ref{eq:suppression_base_def}). The first three parameters are computed by the simulation on-the-fly via the chemical network included in \code{KROME}. Temperature in \code{ramses} is defined from the thermal pressure and the gas density ($\rho = \mu m_H n$, where $\mu$ and $m_H$ are the mean molecular weight and the hydrogen atom mass, respectively) by assuming an equation of state, i.e. $ T = (\gamma -1)\, P_{\rm k}/\rho $, with $\gamma = 5/3$ being the adiabatic index.

To derive the spectrum, in addition to the above mentioned quantities, we need to know the l.o.s. velocity for each cell, namely $v_i = \vec{V_i} \cdot \hat{n}$, where $\vec{V_i}$ is the simulated velocity field of the galaxy and $\hat{n}$ the l.o.s. direction. Having these quantities, we model the contribution of each simulated $i_{th}$ cell to the spectrum as a Gaussian function centred on $v_i$ with a width  $\sigma_{k,t}^i$ and an amplitude equal to the \CII~luminosity ($L_{\rm CII}^{i}$) of that cell. $\sigma_{k,t}^i$ is the broadening of the line for which we account for both the thermal and the turbulent motions as $ \sigma_{k,t}^{i} = \sqrt{(P_{k}^i + P_{t}^i)/\rho_i}$ where $P_{t}^{i}$ is the pressure due to the turbulent motions induced by the kinetic feedback in the simulation. For each velocity bin $v_j$ we compute the integrated line spectrum $f_j = f_j(v_j)$ as:
\begin{equation}
f_j=\sum_{i}\frac{L^{i}_{\rm CII}}{2\sqrt{\pi}\sigma_{k,t}^{i}} e^{-\left[(v_i-v_j)/\sqrt{2}\sigma_{k,t}^{i}\right]^2}
\end{equation}

Having the spectrum as a function of the velocity bin, we define the mean spectral velocity as:
\begin{equation}
\langle v\rangle=\frac{\sum_j v_{j}f_j}{\sum_j f_j},
\end{equation}
which we use to centre the velocities in plotting the spectra.
We compute the Full Width at Half Maximum (FWHM) of the line as the full width at which $68\%$ of the light is contained.; note that in calculating FWHM we do not consider values of $f_j$ lower than $10$ times the peak of the flux.

Note that throughout this paper the \CII~maps are calculated by accounting for the emission of the gas centred on the simulated galaxy that is within a cube with side equal to the field of view (FOV) of the image. Unless noted otherwise, the spectra corresponding to a map are extracted from the same FOV.

\subsection{An example of \CII~surface brightness and spectrum}\label{sec:singlecase}

\begin{figure}
\centering
\includegraphics[width=0.45\textwidth]{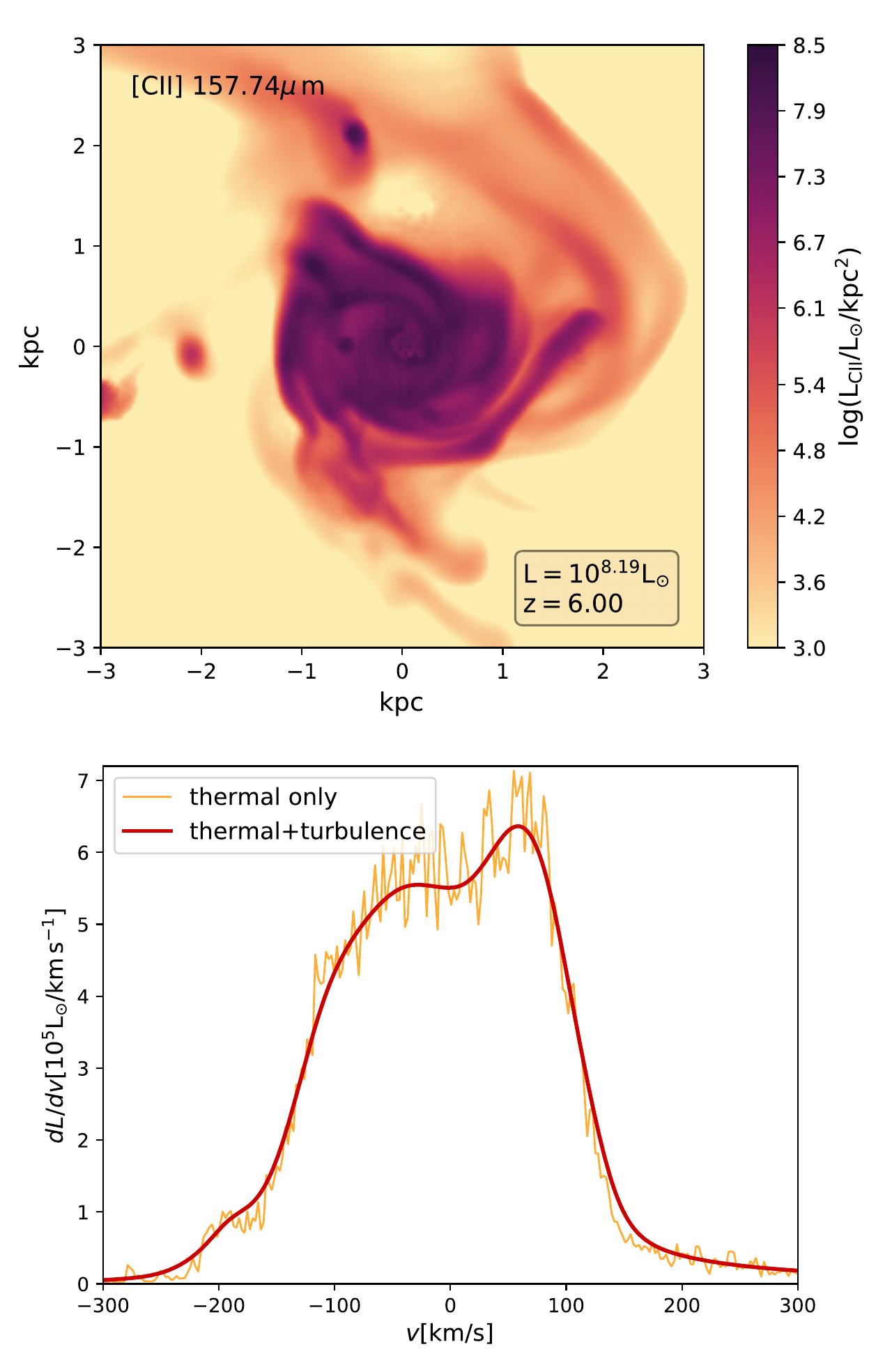}
\caption{{\it Top panel}: Surface brightness of \CII~emission of \althaea~(viewing face-on) at redshift $z=6$. {\it Bottom panel}: The corresponding
synthetic \CII~spectral profile. The spectrum either includes (red line) or does not include (yellow) turbulent broadening; in the latter case, only thermal broadening is taken into account. Turbulent motions smooth out the spectrum by erasing the spiky behaviour and decreasing the line intensity at the peak by $10 \%$. The total spectrum shows two comparable peaks of emission with a relative difference of $< 15 \%$.  }
\label{fig_overview}
\end{figure}

\begin{figure}
\centering
\includegraphics[width=0.5\textwidth]{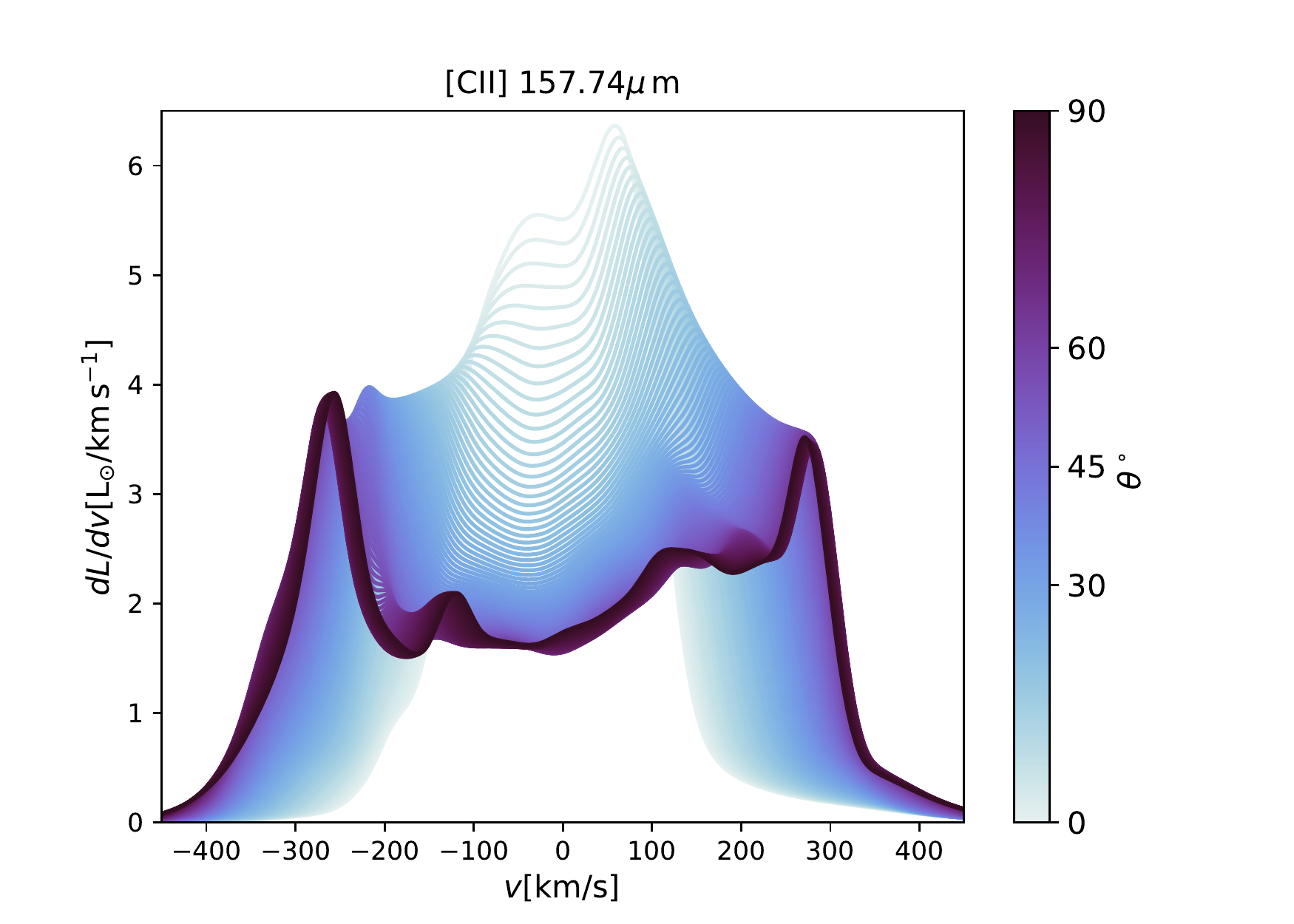}
\caption{Synthetic spectra for \althaea~at $z=6$ (see also Fig. \ref{fig_overview}). Spectra are calculated for $100$ inclinations between face-on ($\theta=0^{\circ}$) and edge-on ($\theta=90^{\circ}$) views.
Changing the inclination clears the signature of rotating disk from the spectral shape, i.e. the two comparable peaks at the edges. From face-on to edge-on the peak amplitude of the line decreases by a factor of $\sim 1.6$.
\label{fig_alth_mult_inclination}}
\end{figure}

We start by discussing the properties of \CII~emission coming from the face-on\footnote{With face-on we mean that we orient the l.o.s. parallel to the eigenvector of the inertia tensor of the gas density distribution with the largest eigenvalue.} view of \althaea~at $z=6$ in a rectangular FOV of size $7\,\rm kpc$ around the centre of the galaxy. In Fig. \ref{fig_overview}, we plot the l.o.s.-integrated surface brightness of the galaxy at this stage. The total \CII~luminosity is  $L_{\rm [CII]}=10^{8.19}\lsun$. The galaxy shows a relatively smooth disk-like structure, whose extent is  $\sim 2 \, \rm{kpc}$ in \CII~emission. At this redshift, this translates to an angular size of $0.34$ arcsec. For these early epochs, there is a clear hint of a broken spiral arm structure. The other interesting feature is the presence of bright clumps of size $\approx 100 \, \rm{pc}$ within the disk.

Also shown in Fig. \ref{fig_overview} is the corresponding synthetic face-on \CII~line spectrum; for comparison, we also present the spectrum in which only thermal broadening is taken into account. The main effect of the inclusion of turbulent motions, self-consistently derived from the simulation in each cell, is to make the line profile smoother by erasing the narrow spikes visible in the thermal-only broadened profile.
As seen in the analytical model (Sec. \ref{sec03}), turbulent motions\footnote{Note that, the turbulence is defined differently in the simulation and the analytical model. In the analytical disk, every motion but the circular ones are treated as turbulence, while in simulation, turbulence is present because of the kinetic feedback. To make an exact comparison, one should fit a disk model to the simulated galaxy and then define the turbulence as it is in the analytical model.} can suppress characteristic features of the spectrum, such as the double-peak profile of a rotating disk.
Note that the maximum of the rotational velocity of the galaxy is of order $\sim 190\, \kms$ (see also in Fig. \ref{fig_vel_struct}), while the level of turbulence for dense gas is of order of $\sigma_t \simeq 30 \, \kms$ \citep{vallini:2018}; thus the effect of micro-turbulence is limited with respect to the range of turbulence explored in the analytical model.
Accounting for turbulent motions in \althaea~decreases the line intensity at the peak by $10 \%$, as it was expected from the analytical model.

As pointed out in Sec. \ref{sec03}, a decreasing inclination can erase the signatures of a disk in the spectra, similar to what happens when increasing turbulent motions. To investigate the situation in our simulated galaxy, we extract \CII~spectra for 100 inclinations between the face-on and the edge-on view of \althaea~disk at $z = 6$ and we plot the result in Fig. \ref{fig_alth_mult_inclination}. Surprisingly, there are two comparable peaks in the spectrum when \althaea~is seen edge-on. This confirms that the gas in the ISM of this galaxy has already undergone ordered rotation at such a high redshift. As expected from our analytical model (see Fig. \ref{fig03}), changing the inclination of the disk washes out the signature of the rotating disk from the spectral profile.
Changing the inclination of the disk from $\theta =0^\circ$ to $\theta=90^{\circ}$, the peak amplitude of the line decreases by a factor of $1.6$. With respect to the analytical disk, spectral profiles contain complicated structures which are due to the asymmetries and clumpy structure of the \CII~ emitting gas. The degeneracy between inclination and turbulent motions is also present in the case of simulated disk but it is more complicated (explored in the analytical model, Fig. \ref{fig03}). Inclining the disk towards face-on not only masks the spectral signature of the disk but also affects the appearance or disappearance of various bumps and structures in the profile.

\subsection{Galaxy evolution traced by \CII}\label{sec:redshift_evolution}

\begin{figure*}
\centering
\includegraphics[width=0.9\textwidth]{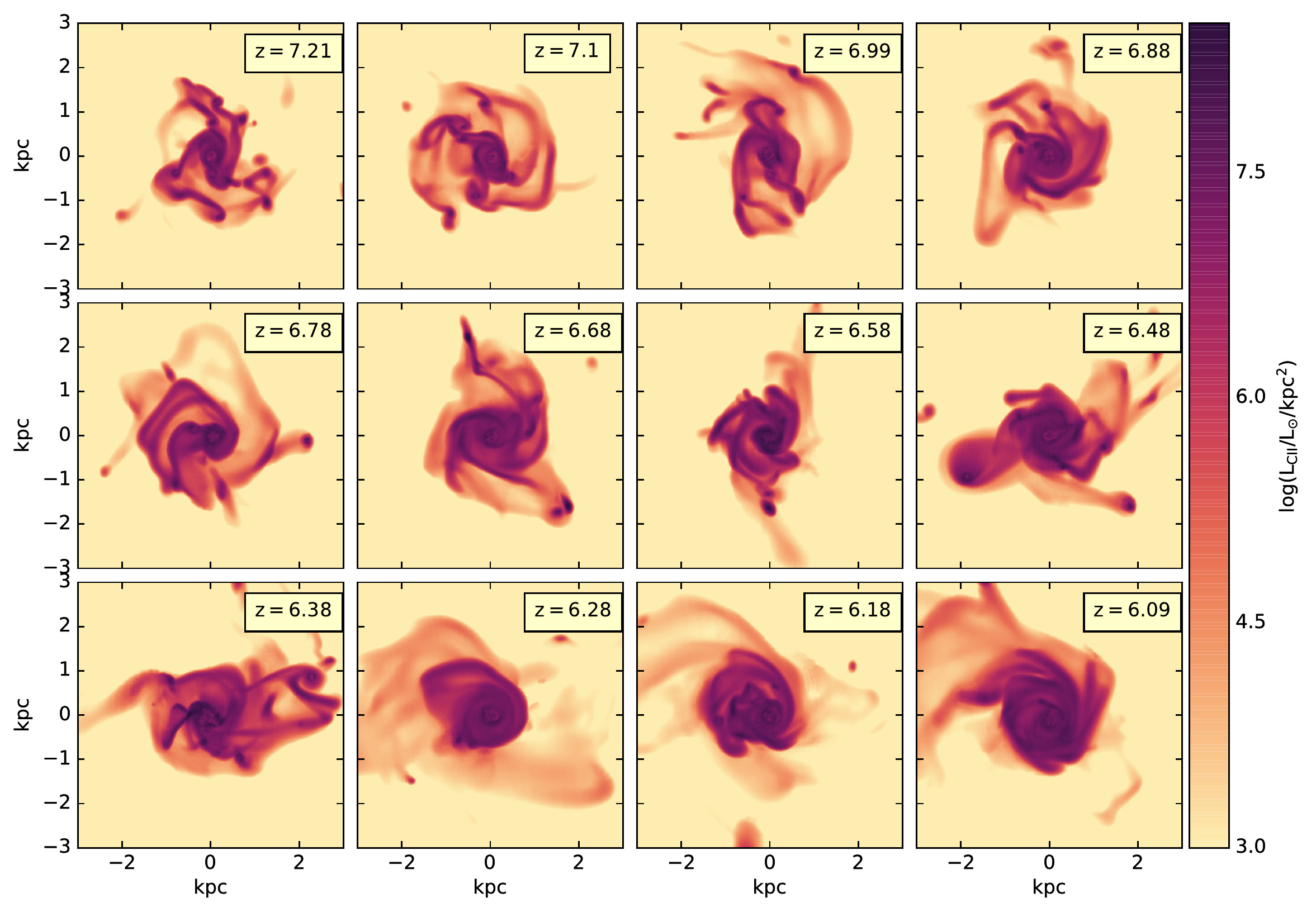}
\caption{\CII~surface brightness of \althaea~during its evolution in redshift range of $6.09<z<7$. Time is increasing from left to right, top to bottom. Neighbouring panels are separated by $\sim 16 \, \rm Myr$. \label{fig06}}
\end{figure*}

\begin{figure*}
\centering
\includegraphics[width=\textwidth]{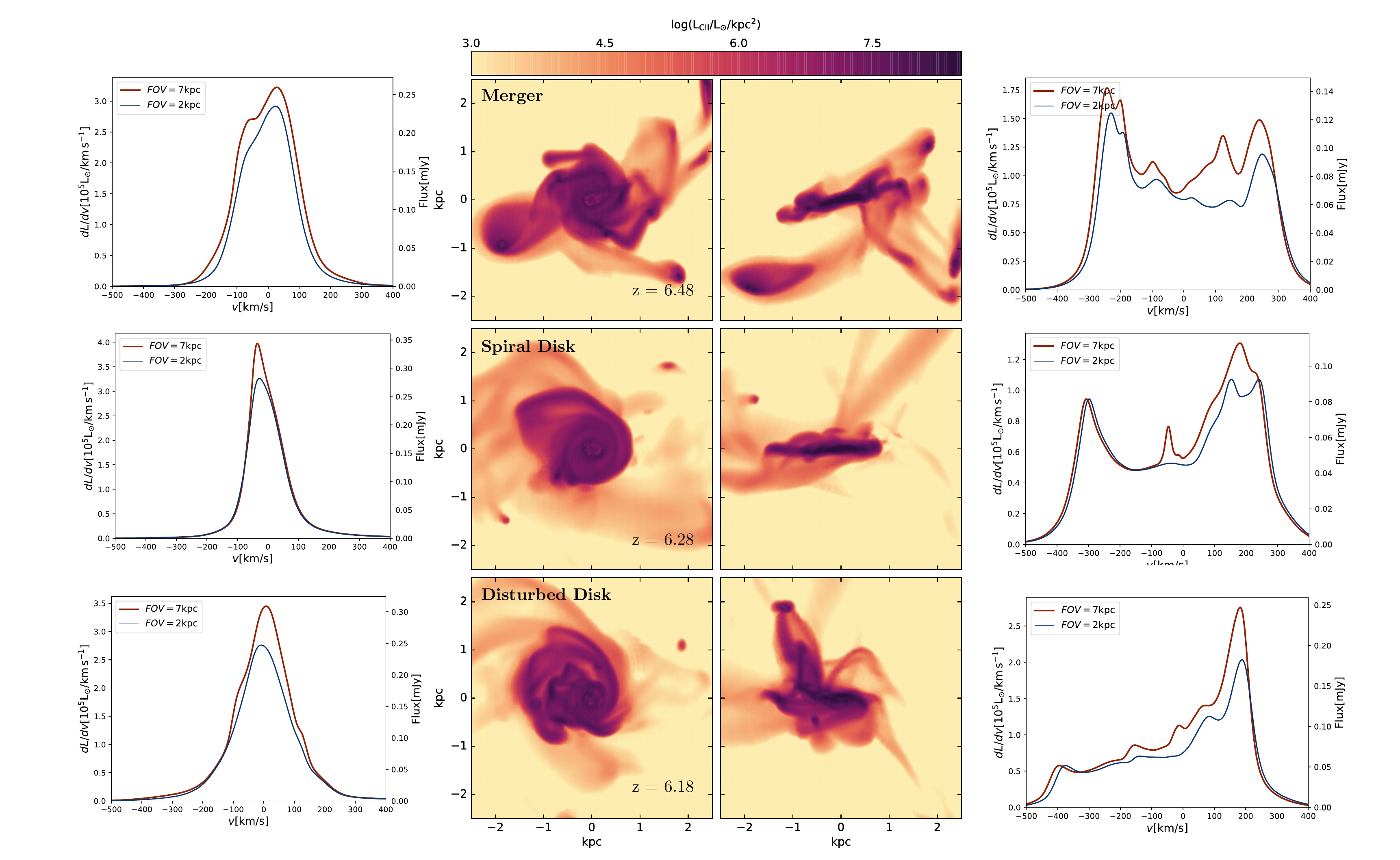}
\caption{Left panel: \CII~ spectra for the face-on inclinations taken from FOV $= 7\,\rm kpc$ (red solid lines) and FOV $= 2\,\rm kpc$ (blue solid lines). Middle panels: Face-on and edge-on emission maps for \althaea~ in different representative stages of evolution: Merger stage (top), Spiral Disk (centre), and Disturbed Disk (bottom). Right panel: \CII~ spectra for the edge-on inclinations taken from FOV $= 7\,\rm kpc$ (red solid lines) and FOV $= 2\,\rm kpc$ (blue solid lines).\label{fig_stages_spectra_maps}}
\end{figure*}

\begin{figure}
\centering
\includegraphics[width=0.5\textwidth]{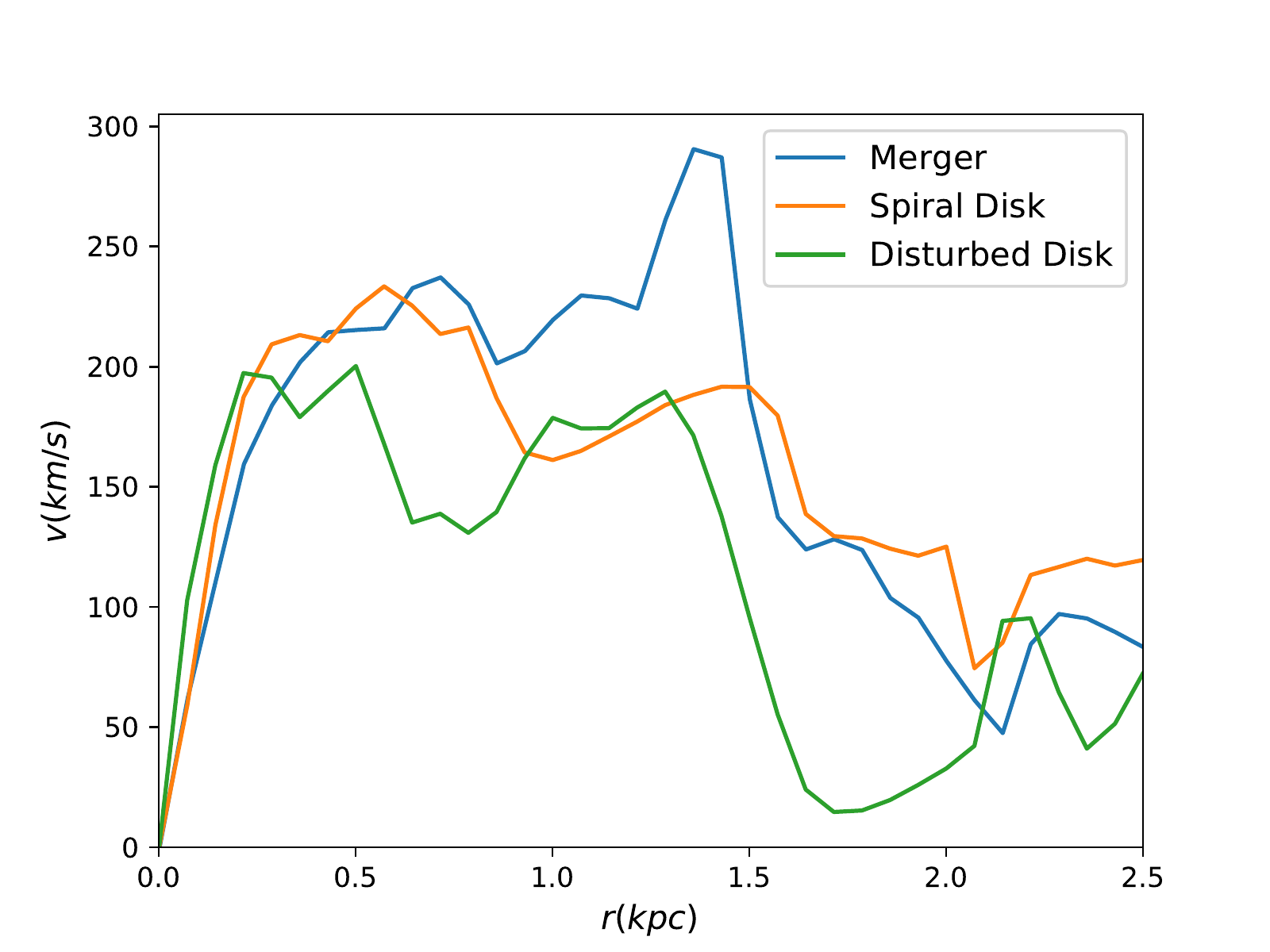}
\caption{Velocity structure of \althaea~in different evolutionary stages (see Fig. \ref{fig_stages_spectra_maps}) in the edge-on view.
\label{fig_vel_struct}
}
\end{figure}

With the tools in hand (emission maps and spectra), now we concentrate on studying the evolution of \althaea~in a redshift range of $6.09<z<7$ (corresponding to a time span of $183\, \rm Myr$) when the system is in a very active assembling phase.
In Fig. \ref{fig06}, we show the face-on emission maps of \althaea~in that redshift range. The time lapse among different panels is $\sim 16 \, \rm{Myr}$ and the images are taken in a FOV of $7\,\rm kpc$.
At the earliest epochs, the galaxy is constituted by a small ($\approx 500 \, \rm pc$) disk surrounded by several emission knots of size $<100 \, \rm pc$, which are feeding the central part through filaments. As time progresses, the disk grows in size and mass in an inside-out fashion, forming a compact core while acquiring mass from the satellites which are progressively disrupted and embedded in the disk.
At $z = 6.48$, a merger event occurs, which is clearly seen in Fig. \ref{fig06}. The merger event dramatically perturbs the quasi-smooth disk structure, resulting in the very irregular and widespread emission seen at $z=6.38$. However, the gravitational potential of the galaxy is able to restore the disk in less than $16 \, \rm Myr$.

Among these stages, we select three particularly interesting stages for further analysis based on their \CII~emission morphology:
\begin{itemize}
  \item[] \textbf{Merger}: at $z=6.48$, when \althaea~experiences a merger event. The satellite in this stage has no stars but is hosted in a dark matter sub-halo that is about to merge with the galaxy. The total \CII~luminosity at this stage is $10^{7.87}L_{\odot}$.
  \item[] \textbf{Spiral Disk}: at $z=6.28$, the ISM of \althaea~has relaxed into a disk which has a spiral arm in one side. The total \CII~luminosity at this stage is $10^{7.71}L_{\odot}$.
  \item[] \textbf{Disturbed Disk}: this stage corresponds to $z=6.18$ in which disk has been vertically disrupted. The total \CII~luminosity at this stage is $10^{7.86}L_{\odot}$.
\end{itemize}
In the two middle panels of Fig. \ref{fig_stages_spectra_maps}, \CII~images for the face-on and edge-on views of the above selected stages are shown. These stages are selected because they have distinct differences in morphology and structure which in principle can evoke differences in the spectral profile of the emission.
Furthermore, we plot the l.o.s. velocity profiles of these stages in Fig. \ref{fig_vel_struct}. These profiles are not monotonic and contain several bumps and peaks. This is an indication of the complex velocity structure of the gas. In the following, we compare these stages of the simulation with each other and also with the cases defined in our analytical model in terms of their spectral profile.

Recall from the analytical model that a double-peak profile is a signature of having a rotating disk in the system while a single peak Gaussian profile can be a signature of either a Disturbed Disk or a face-on view of a rotating disk (Sec. \ref{sec03}).
We apply the spectra diagnostic to the face-on and edge-on views of the above defined stages. As visible in the \CII~images, multiple structures are present in the ISM of these systems, beyond the central $2\, \rm kpc$. To distinguish between the central disk and the environment of the system, we extract the spectra for each of the stages in two FOV sizes, $7\,\rm kpc$ and $2\,\rm kpc$. In the left panels of Fig. \ref{fig_stages_spectra_maps}, these spectra for the face-on view of the stages are plotted, while in the right panels the spectra for the edge-on views are plotted.

The profile of the face-on view of all the stages contain a dominant single peak but they are different in comparison to Smooth Disk and Turbulent dominated Disk defined in the semi-analytical model.

The face-on profile of the Merger stage has a $\rm FWHM=167\,\kms$ and the profile shows two merged peaks located at $v=-100\,\kms$ and $v= 0 \kms$; the major peak is due to the central disk while the addition of \CII~from the starless satellite produces the secondary peak in the profile.
The face-on view of the Spiral Disk with an asymmetric Gaussian shape has a $\rm FWHM=100\,\kms $ and peak flux of $\sim 0.35 \rm mJy$. The asymmetry of the profile reflects the asymmetric kinematics of the \CII~emitting gas. Instead the spectral profile of the Disturbed Disk in face-on view is semi-symmetric but it is wider ($\rm FWHM=143 \,\kms$) in the core because of the extra-planar flows perpendicular to the disk plane; such extra-planar flows can contribute to $\sim 10\%$ of the total signal, as it is analysed in \citet[][]{gallerani:2018outflow}.

In summary, the presence of broken spiral arm,  extra-planar flows and a merging satellite encode spectral signatures as asymmetric Gaussian peaks in the profile, broadening the core of the spectrum and a quite dominant peak very close to the disk's main peak in the face-on spectral profiles respectively.

We perform a similar comparison for the edge-on spectra. The situation for the edge-on profiles is more complicated because the spectra of the simulated stages are very structured.
The edge-on profile of the Merger stage (with peak of $\sim 0.14\,\rm mJy$) contains dominant asymmetric double-peaks (with relative difference of $\sim 18\%$) because of the presence of the central rotating disk. Various bumps are present in the total spectrum and the most prominent one is due to the satellite: its magnitude is comparable to that of the horns of the disk, it is centred around $v= +100\, \kms$ and has a velocity extension of  $\sim 300 \, \kms$. Since this stage shows a clear hint of rotation in the spectrum, it implies that distinguishing systems with close mergers from a rotating system is very difficult using only spectra \citep{Simons+19}.

For the Spiral Disk, the total spectrum has a peak flux of $0.12 \, \rm mJy$ and $\rm FWHM \sim 479\,\kms$; from the spectral shape, there is a clear hint of rotation because of the presence of double peaks in the two edges of the spectrum. The double peaks in the spectral profile are not symmetric as in the analytical model in Smooth Disk. This is because of the asymmetries seen in the disk of \CII~emitting gas (see the right panel of Fig. \ref{fig_stages_spectra_maps}). In addition, there is a quite prominent bump in the core of the spectrum which was not present in the profiles of the analytical disk. The bumps in the spectrum are due to external gas ($> 2\,\rm kpc$) flowing into the disk. In this case, the contribution of the co-planar spiral arm to the edge-on spectrum becomes more prominent making the high-velocity tails.

As it is expected from the analytical model, the edge-on profile of the simulated Disturbed Disk does not have rotating double peaks. However, instead of having a smooth single Gaussian profile, there is an asymmetric Gaussian profile(centred on $\rm v = 200\kms$) including multiple peaks in the long skewed tail. There is a relative difference of $\sim 80\%$ between the main peak of the spectrum and the lowest bump in the tail. The presence of extra-planar flows suppresses the blue part of the spectrum masking the signature of rotating disk which was present in the Spiral Disk profile. Recall that Disturbed Disk stage is just $16\, \rm Myr$ after the Spiral Disk stage in the evolution of \althaea.

The spectra for the simulated galaxy are very structured and complicated. To properly interpret the component analysis, it is required to apply full dynamical studies and extract the spectra for different velocity channels of the system. This is beyond the scope of the present paper and is left for future studies.

\section{Observational Implications}\label{sec05}

\begin{figure}
\includegraphics[width=0.485\textwidth]{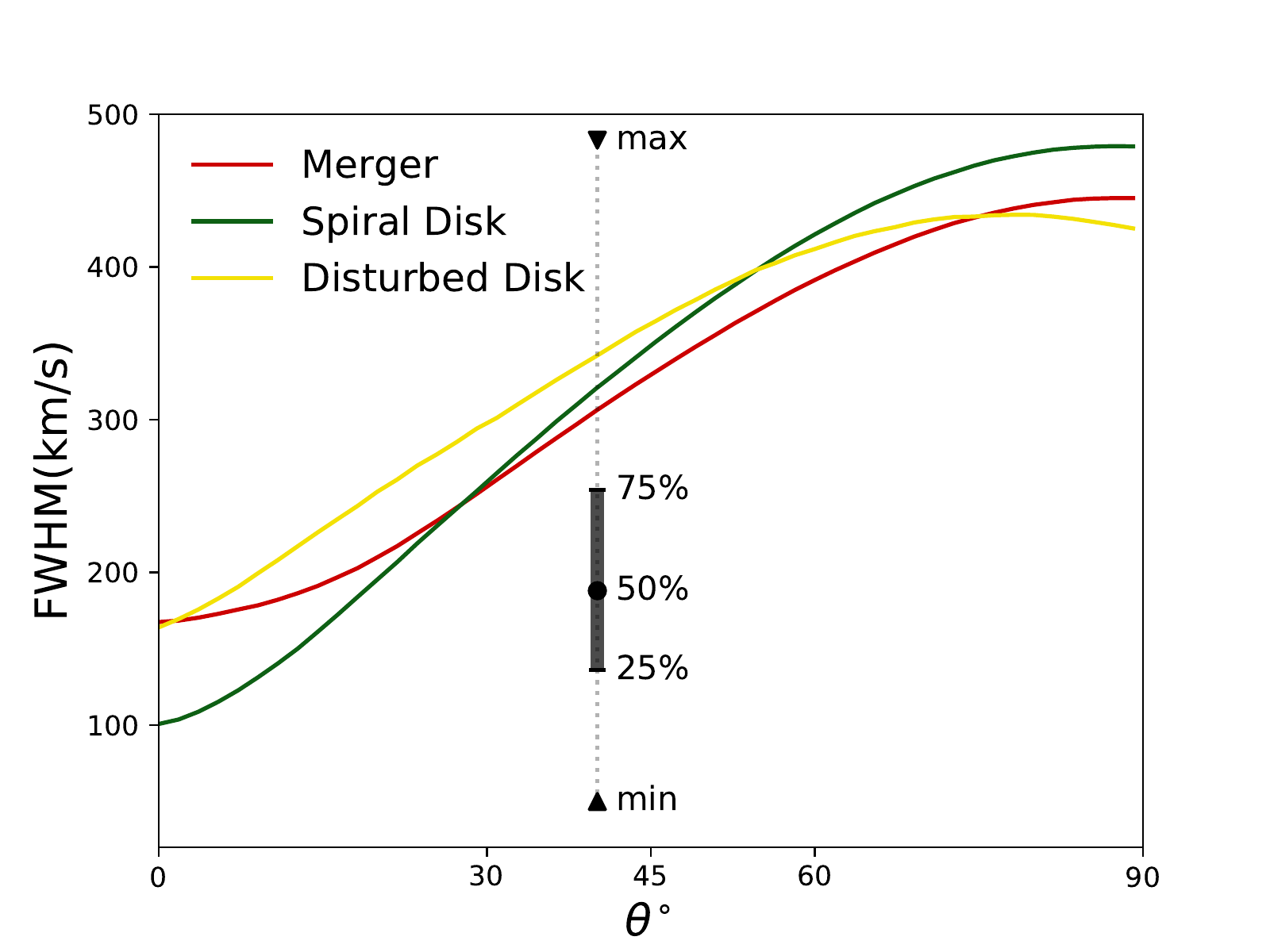}
\caption{
The FWHM of \althaea~spectra as a function of inclination. Each line corresponds to a different evolutionary stage (see Fig. \ref{fig_stages_spectra_maps}).
As a reference, we over-plot the statistical properties of the FWHM from observed in high-redshift galaxies (see Table \ref{tab-obs}).
\label{fig_fwhm}}
\end{figure}
\begin{figure*}
\includegraphics[width=0.33\textwidth]{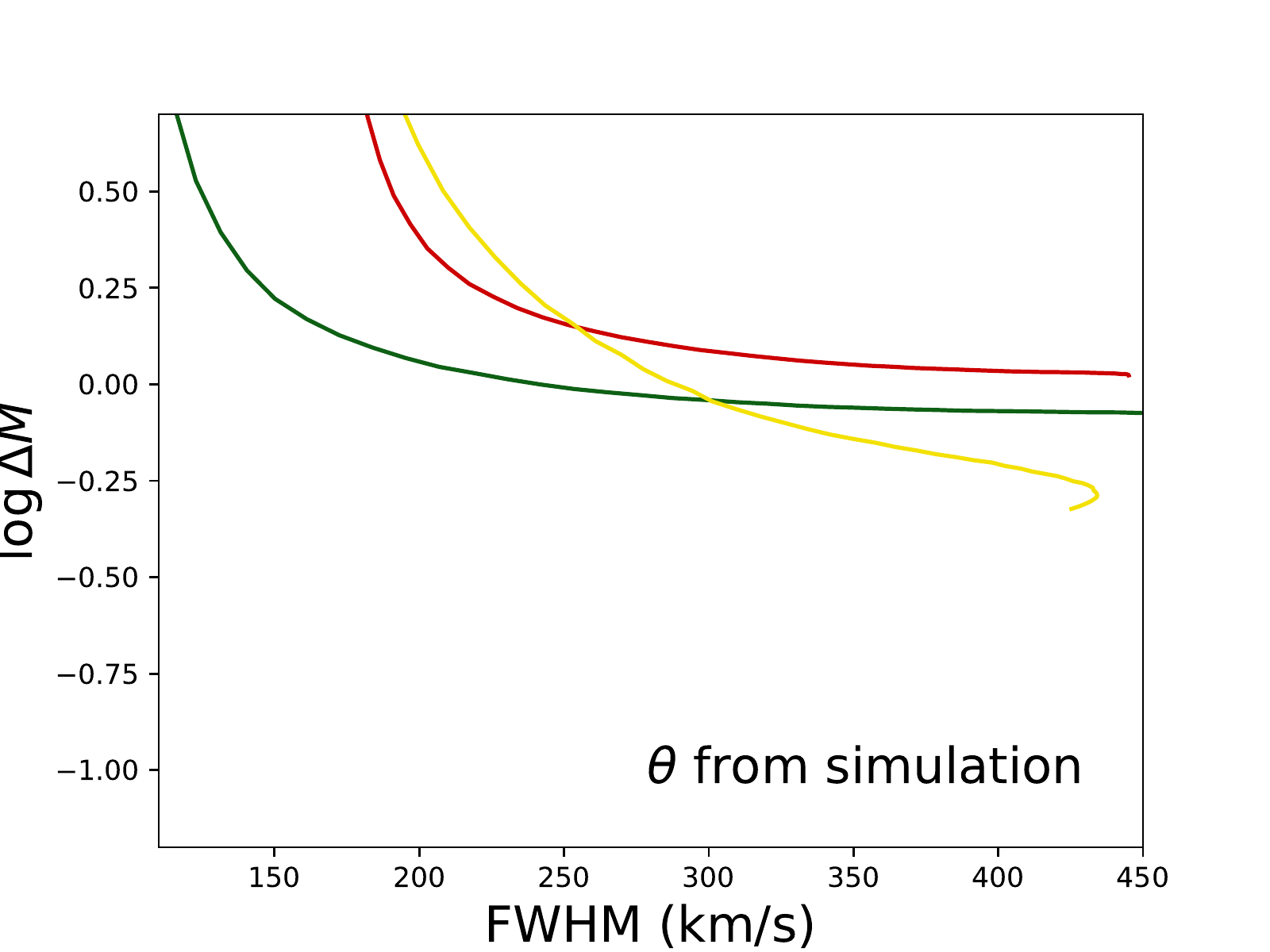}
\includegraphics[width=0.33\textwidth]{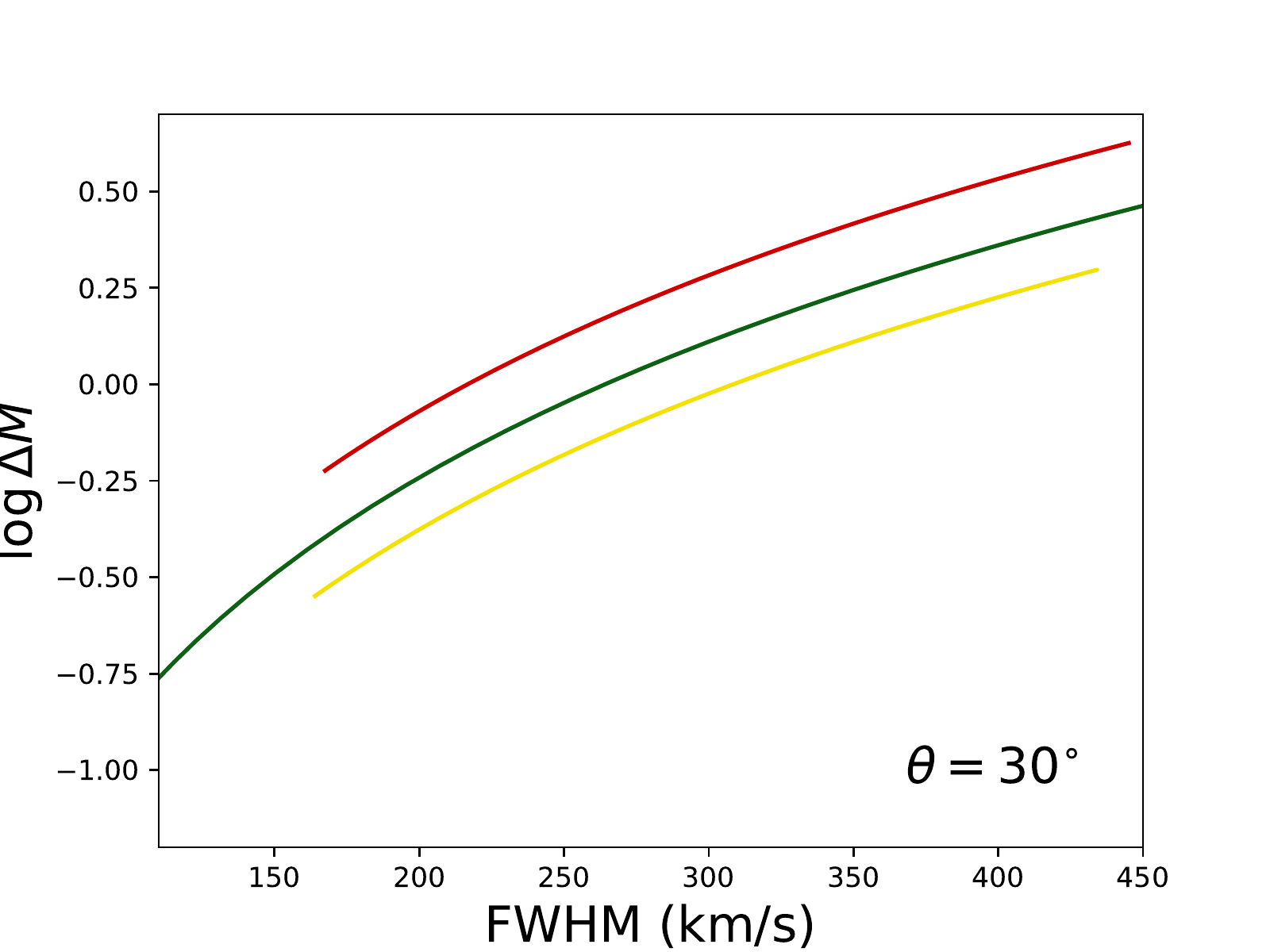}
\includegraphics[width=0.33\textwidth]{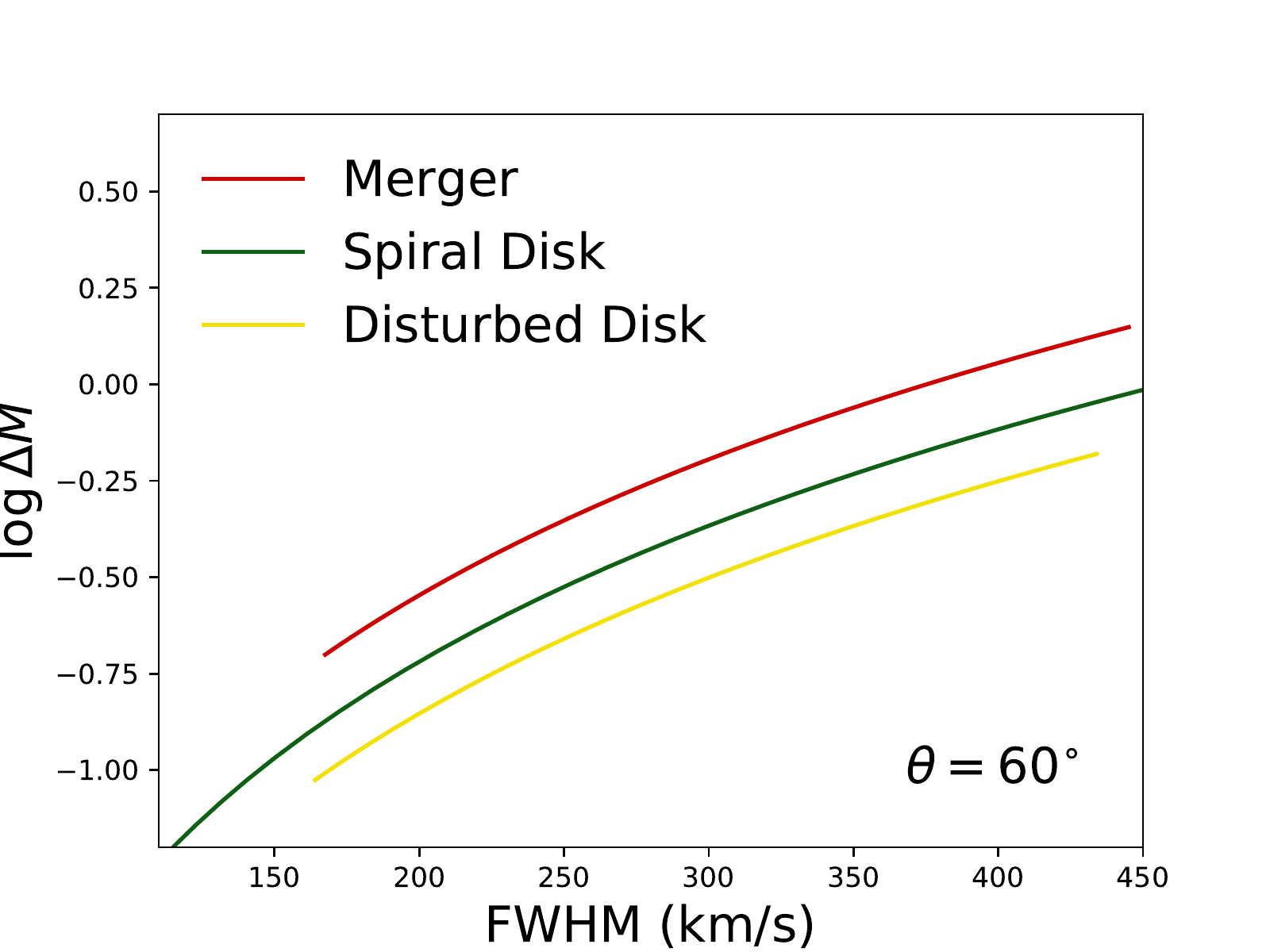}
\caption{
Analysis of the error in the dynamical mass determination using empirical estimates \ref{eqn_mdyn}. The mass-error-function ($\Delta M$, eq. \ref{eq_delta_m}) is plotted as a function of the FWHM for known inclination, as calculated in the simulation (left panel), and by assuming a fixed inclination of $30^{\circ}$ (central panel) and $60^{\circ}$ (right panel). Different lines indicate the three selected evolutionary stages (see Fig. \ref{fig_stages_spectra_maps}).
\label{fig_delta_m_vs_fwhm}}
\end{figure*}

\begin{figure}
\includegraphics[width=0.485\textwidth]{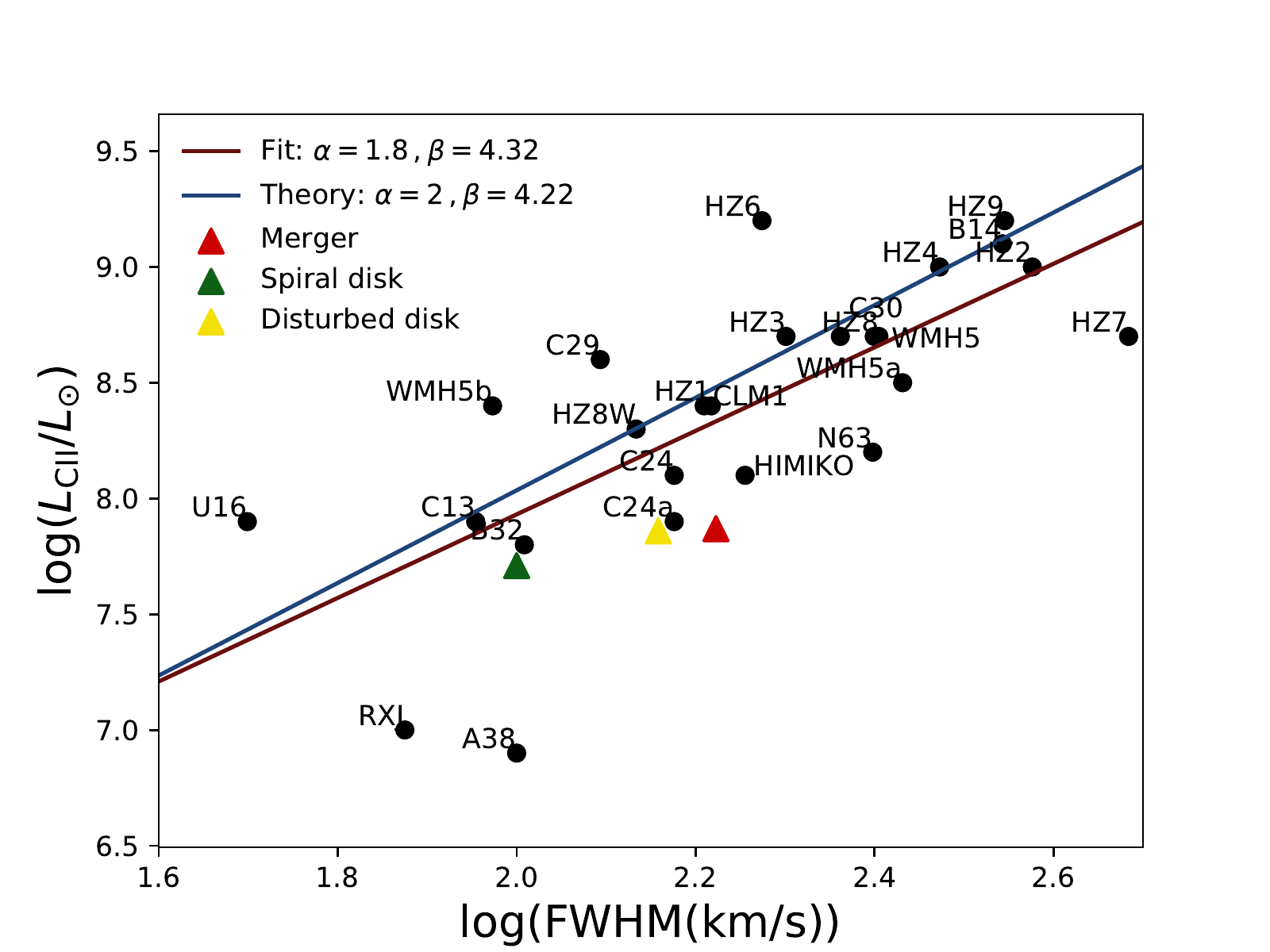}
\caption{
Correlation between $L_{\rm CII}$ and FWHM for the sample of observed high-$z$ galaxies (black dots, Tab. \ref{tab-obs}). The red line indicates the fit to the data (functional form and parameters in eq.s \ref{eq_system_fit}). The blue line indicates the approximate relation given in eq. \ref{eqn_TF_approx} assuming $R = 1\,\rm kpc$ and $\sin{\theta} = 0.5$.
\label{fig_lcii_fwhm}
}
\end{figure}
\begin{figure*}
\centering
\includegraphics{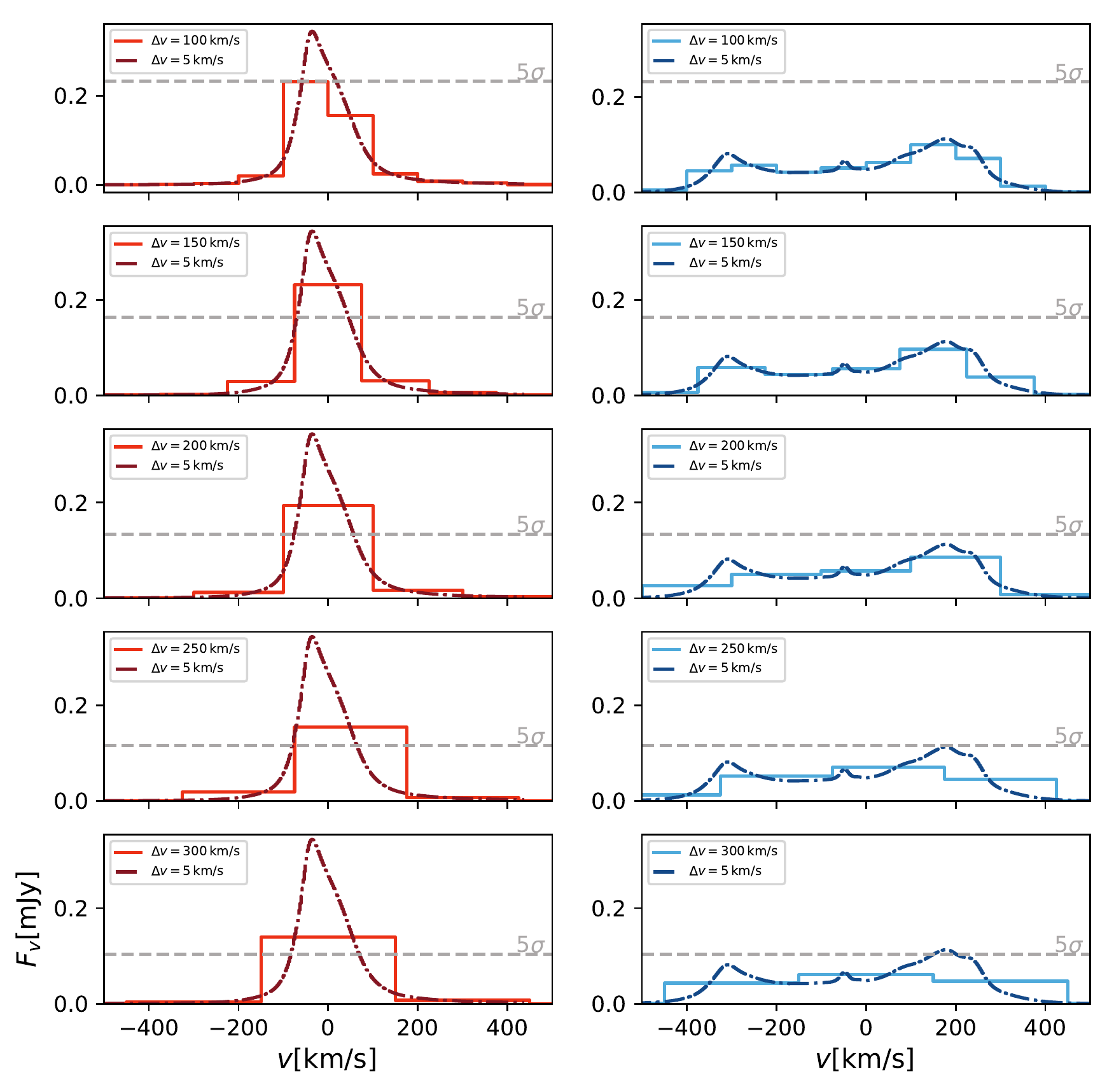}
\caption{
\label{fig:rebinning}
Comparison of the \CII~observability for face-on orientations (left panels) vs edge-on (right panels) ones when \althaea~has $\log (L_{\rm CII}/\lsun) = 7.7$, at $z=6.28$ (Spiral Disk). From top to bottom the spectrum is re-binned with an increasing channel width ($\Delta v$). As a reference, the $\Delta v = 5\kms$ case is reported in all panels. The dashed horizontal line corresponds to $5\sigma$ noise level, that is calculated by assuming a 10 hours ALMA observation.
}
\end{figure*}

\begin{table*}
\centering
\caption{Sample of high-$z$ galaxies probed by \CII~line.
\label{tab-obs}
}
\begin{tabular}[t]{lccccc}
\hline
Target Name & ID & $z$ & $\log(L_{\CII}/\lsun)$ & FWHM$/\rm km\,s^{-1}$& Reference\\
\hline
UDS16291 &U16 & 6.64 & 7.9 & 50& \citet{Pentericci+16}~\\
RXJ1347:1216 &RXJ& 6.77 & 7 & 75&  \citet{Bradac+17}~\\
COSMOS13679 &C13 & 7.15 & 7.9 & 90& \citet{Pentericci+16}~ \\
WMH5b&WMH5b & 6.07 & 8.4 & 94& \citet{Jones+17,Willot+15}~ \\
A385-5.1 &A38 & 6.03 & 6.9 & 100& \citet{Knudsen+16}~ \\
BDF3299 &B32 & 7.15 & 7.8 & 102& \citet{Maiolino+15,Carniani+17} \\
COS-2987030247 &C29 & 6.81 & 8.6 & 124&\citet{Smit+18} \\
HZ8w&HZ8W & 5.15 & 8.3 & 136& \citet{Capak+15} \\
COSMOS24108a &C24a& 6.63 & 7.9 & 150&\citet{Pentericci+16} \\
COSMOS24108&C24 & 6.63 & 8.1 & 150& \citet{Pentericci+16} \\
BDF2203&B22 & 6.12 & 8.1 & 150& \citet{Carniani+18clumps} \\
CLM1&CLM1 & 6.17 & 8.4 & 162&\citet{Willot+15} \\
HZ1&HZ1 & 5.69 & 8.4 & 165&\citet{Capak+15} \\
HIMIKO&HIMIKO & 6.60 & 8.1 & 180&\citet{Ouchi+13,Carniani+18himiko}  \\
HZ6&HZ6 & 5.29 & 9.2 & 188& \citet{Capak+15} \\
HZ3&HZ3 & 5.54 & 8.7 & 200&\citet{Capak+15} \\
COS-3018555981 &C30 & 6.85 & 8.7 & 230 &\citet{Smit+18}\\
NTTDF6345&N63 & 6.70 & 8.2 & 250&\citet{Pentericci+16} \\
WMH5&WMH5 & 6.07 & 8.7 & 251& \citet{Jones+17,Willot+15} \\
HZ8&HZ8 & 5.15 & 8.7 & 254&\citet{Capak+15}\\
WMH5a&WMH5a & 6.07 & 8.5 & 270& \citet{Jones+17,Willot+15} \\
HZ4&HZ4 & 5.54 & 9.0 & 297&\citet{Capak+15} \\
B14-65666&B14 & 7.15 & 9.1 & 349&\citet{Hashimoto+18} \\
HZ9&HZ9 & 5.54 & 9.2 & 351&\citet{Capak+15} \\
HZ2&HZ2 & 5.66 & 9.0 & 377&\citet{Capak+15} \\
HZ7&HZ7 & 5.25 & 8.7 & 483&\citet{Capak+15} \\
\hline
\end{tabular}
\end{table*}%

Investigating the evolution of \althaea, we have seen how the structural and kinematical differences result in various spectral profiles which depend on morphological properties, and inclination of the galaxy (Fig. \ref{fig_alth_mult_inclination}). In this Section, we analyse the implications of these results from an observational point of view.

For the synthetic spectra, we use the three different stages of \althaea~discussed in Sec. \ref{sec:redshift_evolution}.
Our results are compared with observations of a sample of $5.2<z<7.1$ galaxies for which the spectra of \CII~line have been obtained with ALMA \citep{Ouchi+13,Wang+2013,Capak+15,Pentericci+16,Jones+17,Carniani+17}. For reference, these objects are listed in Table \ref{tab-obs}, along with their redshift, total \CII~luminosity ($L_{\rm CII}$) and FWHM of the \CII~line.

\subsection{Dynamical mass estimates}\label{sec_dyn_masses}

By assuming a rotating disk geometry (with radius $R$) for the \CII~emitting gas, the dynamical mass can be estimated as:
\be \label{eq_dyn_mass_gen}
M_{\rm dyn} = \frac{v_c^2 R}{G}\,.
\ee
From a \CII~spectrum obtained with a high signal to noise ratio and a good sampling of the velocity channels one can estimate $v_c$  from the FWHM of the line using the following expression:
\be\label{eqn_fwhm_vc}
{\rm FWHM} = \gamma v_c \sin{\theta}\,,
\ee
where $\gamma$ is a factor of order of unity that depends on geometry, line profile, and turbulence. Different values have been assumed in the literature for $\gamma$: for example, \citet{Capak+15} assumed $\gamma = 1.32$.
Using eq.s \ref{eq_dyn_mass_gen} and \ref{eqn_fwhm_vc}, the general expression for the dynamical mass is:
\be \label{eqn_mdyn}
\rm M_{\rm dyn}^{est} = 2.35 \times 10^{9}\msolar \left(\frac{1}{\gamma^2 \sin^2{\theta}}\right)\left(\frac{\rm FWHM}{100\,\kms}\right)^{2}\left(\frac{R}{\rm kpc}\right)\,.
\ee

Before discussing the mass estimates, let us consider the FWHM of the spectra. We plot them as a function of inclination in Fig. \ref{fig_fwhm}. In general, the FWHM in \althaea~is an increasing function of inclination and varies from a minimum of $100 \,\kms$ in the face-on case to a maximum of $480\,\kms$ for the edge-on case.
In the same Figure we compare the simulated FWHM with the one inferred from observations of high-$z$ galaxies (Tab. \ref{tab-obs}).
The bulk of the observed spectra have a \CII~line FWHM around $180 \,\kms$ that is compatible with that found from \althaea~seen face-on. Note that \althaea~has a dynamical mass $M_{\rm dyn}\simeq 10^{10}\,\msun$, while the dynamical masses of the observed galaxies range from $10^9 \msun$ to $10^{11}\msun$ \citep{Capak+15}.

We are interested in assessing the reliability of the dynamical mass estimates obtained from eq. \ref{eqn_mdyn} as a function of \CII~line FWHM. The radius of the disk is computed from the \CII~image as $R \simeq  r_{80}$, i.e. the effective radius of of the system containing $80\%$ of the total \CII~luminosity. For the three aforementioned stages of \althaea, $R\simeq 1\, \rm kpc$.
It is convenient to define the \quotes{mass-error-function}, i.e.
\be\label{eq_delta_m}
\Delta M \equiv M^{\rm est}_{\rm dyn}/M_{\rm dyn}\,,
\ee
that parametrises the error in the mass estimates using eq. \ref{eqn_mdyn}, that depends on $\gamma$. We calculate $\gamma$ from our simulation depending on the stage of the evolution. The value of $\gamma$ for the Spiral Disk, Disturbed Disk and the merger stage is $1.78$, $2.03$ and $1.52$, respectively.

In Fig. \ref{fig_delta_m_vs_fwhm}, we plot $\Delta M$ as a function of the FWHM for \althaea.
In the left panel estimates are performed by using the information on the inclination obtained from the simulation. The minimum of $\Delta M =0.4$ is found for large FWHM ($\ge 350\, \kms$). At low FWHM $\Delta M$ becomes very large as $M^{\rm est}_{\rm dyn}\propto 1/\sin^2\theta$. In all cases we find $\Delta M> 1$ for low FWHM and $\Delta M< 1$ at high FWHM.
This means that by using eq. \ref{eqn_mdyn} we tend to underestimate (overestimate) the dynamical mass at high (low) FWHM, or, equivalently, inclinations (see the left panel of Fig. \ref{fig_delta_m_vs_fwhm}).

It is interesting to calculate the mass-error-function for fixed inclinations, $\theta=30^{\circ}$ and $\theta=60^{\circ}$. These two values are generally assumed when $\theta$ cannot be directly determined from observations. This can happen when the spatial resolution does not allow us to constrain the inclination, as in \citet{Capak+15}, that calculate the dynamical masses by assuming $\sin{\theta}= 0.45-1$.
Results are shown in the central and right panels of Fig. \ref{fig_delta_m_vs_fwhm}. For $\theta = 60^{\circ}$, $\Delta M<1$ except for the high inclinations of the merger stage, while for $\theta = 30^{\circ}$ the dynamical mass is typically overestimated, up to a factor $\simeq 4$.
The error of the estimate is comparable with the one reported for the sample of \citet{Capak+15}, where the authors concluded that at $z>5$ the dynamical masses are typically a factor of $\sim 3$ greater than the stellar masses. This should be confronted with the analogous factor of $1.2-1.7$ measured at $z\sim 1-3$ \citep{schreiber+09}.

The mass estimates eq. \ref{eqn_mdyn} is based on the assumption that the galaxy has a smooth disk. However, our simulations show that high-$z$ galaxies have more complex dynamical structures which result in correspondingly complex spectra. As observations are progressively becoming more precise, a better modelling of kinematics and velocity structure of the galaxies is required \citep[e.g.][]{BBarolo+15}.

\subsection{Tully-Fisher relation for high-$z$ galaxies}

In Fig. \ref{fig_lcii_fwhm}, we plot the observed $L_{\rm [CII]}$ -- FWHM relation for the high-$z$ galaxy sample in  Tab. \ref{tab-obs}. The best-fit to the data is
\begin{subequations}\label{eq_system_fit}
\be\label{eqn_fit}
\log(L_{\rm[CII]}/L_{\odot})=\alpha \log(\rm FWHM/\kms) + \beta\,,
\ee
with
\begin{align}
\alpha&=1.80\pm 0.35\,,\\
\beta &=4.32\pm 0.78\,.
\end{align}
\end{subequations}
The Pearson coefficient is $\simeq 0.74$, suggesting a statistically reliable correlation between these two parameters. The three stages of \althaea~(viewed face-on) are shown as triangles in this plot. They fall within $1\sigma$ from the best-fit curve.

Such relation resembles the \citet{Tully+77} relation. Its existence is not surprising because of the link between $L_{\rm[CII]}$ and the dynamical mass.
As a rough estimate (see \citep{pallottini:2017dahlia} for an extensive discussion), we can assume a constant ratio between the total \CII~luminosity and the gas mass in a high-z galaxy; thus we can write
\be\label{eq_eta}
\kappa = \frac{L_{\rm[CII]}}{M_{g}}=\frac{L_{\rm[CII]}}{f_g M_{b}}\,,
\ee
where $f_g$ is the gas fraction of the baryonic mass ($M_b$). Using eq. \ref{eqn_mdyn} and defining $f_{\rm DM}=M_{\rm DM}/M_{b}$ as the ratio between dark matter and baryonic mass, the relation between $L_{\rm [CII]}$ and the FWHM of the line reads as
\be \label{eqn_TF}
L_{\rm [CII]} = \frac{\kappa R}{G}\left(\frac{f_g}{1+f_{\rm DM}}\right)\left(\frac{\rm FWHM}{\gamma \sin{\theta}}\right)^2\,.
\ee
Interestingly, this simple analytical expression is consistent with the empirical relation (eq. \ref{eq_system_fit}). It is convenient to express eq. \ref{eqn_TF} in terms of typical values found in high-$z$ galaxies.
Roughly, from our model we expect $\kappa = 0.1 {\rm L}_{\odot}/{\rm M}_{\odot}$ (eq. \ref{eqn:ciiemissivity}, see also \citealt{pallottini:2017althaea}), $\gamma \simeq 1.7$, $f_{g}\simeq 0.5$, and $f_{\rm DM}\simeq 0.5$; thus eq. \ref{eqn_TF} can be written as
\be \label{eqn_TF_approx}
L_{\rm [CII]} \simeq
5.4 \times 10^{7}{\rm L}_{\odot} \left(\frac{1}{\sin^2{\theta}}\right)\left(\frac{\rm FWHM}{100\,\kms}\right)^{2}\left(\frac{R}{\rm kpc}\right)\,.
\ee
Further, fixing $R = 1 \, \rm kpc$ and $\sin{\theta} = 0.5$ we can express $\log L_{\rm CII}$ vs FWHM as in \ref{eqn_fit} with parameters
\begin{subequations}
\begin{align}\label{tf_like_theory}
\alpha&= 2 \\
\beta &= 4.22\,
\end{align}
\end{subequations}
which is within $1\sigma$ from the fit (eq. \ref{eq_system_fit}).

As a final remark, we note that in Fig. \ref{fig_lcii_fwhm}, there is a lack of data in both the low FWHM-high $L_{\rm [CII]}$ and the high FWHM-low $L_{\rm [CII]}$ regions.
While the first occurence is physically motivated (it is unlikely that low mass galaxies have large luminosities), the second one might arise from an observational bias. In fact, as \CII~is optically thin, its luminosity is constant with inclination. As a consequence, as the FWHM increases, the peak flux might drop below the detection threshold. We investigate this issue in the next Section.

\subsection{Observations of edge-on vs face-on galaxies}\label{sec_edgeon_vs_faceon_bias}

We now check the detectability of \CII~line for face-on and edge-one inclinations by performing mock ALMA observability simulations.
We select the Spiral Disk evolutionary stage, i.e. when \althaea~has luminosity $\log (L_{\rm CII}/\lsun) = 7.7$, similar to the one inferred for BD3299 \citep{Maiolino+15,Carniani+17}.
As for BD3299 observation \citep{Carniani+17}, we assume a $10$ hours integration time with ALMA. We consider the edge-on and face-on inclinations and we re-bin the spectra with channel width in the range $100\leq \Delta v\leq 300$, i.e. the typical one used when searching for lines in normal star forming galaxies (${\rm SFR}\lsim 100 \msun /\rm yr$).

The results of such analysis are shown in Fig. \ref{fig:rebinning} where we also plot the $5\sigma$ noise level for some selected values of $\Delta v$.
The face-on case is detected at $>5\sigma$ for all considered $\Delta v$, thus yielding a ${\rm FWHM}\sim 100\,\kms$ which is very similar to what is reported for BDF3299 in \citet{Carniani+17}.
However, the edge-on case with a larger intrinsic {$\rm FWHM= 479\,\kms$} would be always undetected. Stated differently, the large l.o.s. velocities smear out the spectrum, making the detection more challenging if the galaxy is seen edge-on.
This suggests that some of the non-detections reported at high-$z$ might be due to inclination effects when the target is close to edge-on.
Note that here we are assuming that no beam smearing effects are in place, that is equivalent to assume that we marginally resolve the flux from the galaxy.
This interpretation must be substantiated in a future work with better quantifying channel noise and spatial correlations of the ALMA beam.

\section{Summary and Conclusions}\label{sec06}

We have studied the structural and kinematical properties of galaxies in the Epoch of Reionization ($z\geq 6$) as traced by the spectral profile of the \CII~emission line. The emission is computed from an analytical model accounting for gas cooling via the \CII~line \citep{dalgarno:1972,wolfire+95,vallini:2013}, and it includes CMB suppression of the line intensity \citep{DaCunha2013,Pallottini+15,vallini:2015}.

First, we have applied our model to an idealised rotating disk galaxy, in order to investigate the effect of disk inclination ($\theta$) and turbulent velocities ($v_t$) on the line profile. From this controlled environment, we have found that both large turbulent motions ($v_t / \bar{v}_c=\zeta>0.5$, where $\bar{v}_c \simeq 75\, \kms $ is the galaxy circular velocity) and inclination angles $\theta <75^\circ$ erase the double-peak line profile, expected from a rotating disk galaxy.
In particular, we find that the peak flux of \CII~emission for face-on ($\theta=0^\circ$) can be a factor $\sim 4$ higher than in the edge-on view ($\theta=90^\circ$).
Next, we have used zoom-in cosmological simulations of a prototypical Lyman break galaxy \citep[\quotes{Alth{\ae}a}, ][]{pallottini:2017althaea} to analyse the \CII~emission properties during its evolution in the redshift range $6\lesssim z\lesssim 7$. Information on velocities, thermal, and turbulent motions included in the simulation, enabled us to build the \CII~surface brightness maps of \althaea~and the synthetic spectra.
At $z=6.0$, Alth{\ae}a has a total \CII~luminosity $L_{\CII}=10^{8.19}L_{\odot}$; this value accounts for a factor $\sim 2$ suppression due to the CMB (see Fig. \ref{fig05}). At this epoch and viewed face-on, the \CII~emission map shows a smooth, disk-like structure with an extent of $\sim 2\, \rm kpc$, on top of which are superimposed clumps with typical sizes of $\sim 100 \, \rm pc$.
From the analysis of the \CII~line profile, we find that the effect of turbulent motions is to smooth out the spectrum by broadening the thermal profiles and to decrease the peak line intensity by $\sim 10\%$. The degeneracy between turbulent motions and inclination is also present in the spectra of \althaea, that has a $\zeta\simeq 0.15$. The edge-on spectral profile of this stage is indicative of a rotating disk, i.e. it shows a double peak profile. Decreasing the inclination progressively washes out the disk signature from the profile and increases the peak flux by a factor of $\simeq 1.6$.

Studying the morphology of Alth{\ae}a in the redshift range $6\leq z\leq 7$, we have identified three main evolutionary stages with distinct spectral signatures: I) Merger, II) Spiral Disk, and III) Disturbed Disk. The irregular and choppy structure of the l.o.s. velocity profiles resulting from the simulations (see Fig. \ref{fig_vel_struct}) translates into more structured \CII~line profiles with respect to the analytical model. Comparing the synthetic spectra for different stages of \althaea~with the ones from the analytical model, we identify the spectral signatures of  merger events, spiral arms and extra-planar flows in the respective stage both in the face-on and edge-on profiles. The main signatures are summarised as follows:

\begin{itemize}
\item[]{\bf Merging Satellites}: the face-on profile of the merger stage of \althaea~has a peak flux of $\sim 0.27\, \rm mJy$, with a second peak in the blue part centred on $v=-100\,\kms$. The major peak of the spectrum is due to the central disk, while the second peak is produced by the starless satellite. In the edge-on case, the spectrum shows an asymmetric double peak along with multiple peaks in the core due to co-rotating clumps. The signature of the merging satellite is visible as a broad peak (with spectral extent of $300\,\kms$) in the red side of the double peak profile (centred around $v=+100\, \kms$).
\item[]{\bf Spiral arms} manifest in the asymmetric Gaussian profile of the face-on spectrum of the Spiral Disk stage. In the edge-on view, the signature of spiral arms is contained in the asymmetry of the double-peak profile corresponding to the rotating disk.
\item[]{\bf Extra-planar flows}: the \CII~spectrum for the face-on view of Disturbed Disk stage features a quasi-symmetric Gaussian profile which has a broader core and more prominent wings compared to the Spiral Disk. Instead, in the edge-on view, extra-planar flows tend to erase the blue peak of the  line profile, hence masking the rotating disk characteristic feature.
\end{itemize}

Finally, we have discussed the observational implications of our analysis by comparing them to \CII~observations of high-$z$ galaxies (see Tab. \ref{tab-obs}). The bulk of the observed spectra have $\rm FWHM\sim 180\, \kms$, that is compatible with face-on spectra of Alth{\ae}a. Our key results are the followings:

\begin{itemize}
\item[] {\bf Dynamical mass estimates}: we derived a generalised form of the dynamical mass vs. \CII-line FWHM relation (eq. \ref{eqn_mdyn}) which depends on the dynamical state of the galaxy. If precise information on the galaxy inclination is available, the returned mass estimate is accurate within a factor $\sim 2$. If the inclination is not constrained, the error increases up to a factor of $\simeq 4$. These errors are due to the fact that high-$z$ galaxies have a complex dynamical structure and the assumption of a smooth disk used in the derivation of eq. \ref{eqn_mdyn} is not fully valid.
\item[] {\bf Tully-Fisher relation}: we find a correlation between the $L_{\rm[CII]}$ and FWHM of the \CII~line by fitting the values for the sample of high-$z$ galaxies, i.e. $L_{\rm[CII]}\propto (FWHM)^{1.80\pm 0.35}$ (eq.s \ref{eq_system_fit}). This can be understood from simple physical arguments that are embedded in the relation given in eq. \ref{eqn_TF}. By fixing the inclination and radius of the galaxy, we find that such approximate theoretical expression (eq. \ref{eqn_TF_approx}) is consistent with the empirical relation.
\item[] {\bf Inclination and detectability}: we have performed mock ALMA simulations to check the detectability of \CII~line for face-on and edge-on views. We consider a fixed integration time (10 hr) and rebin the spectra of the Spiral Disk stage with channel width in the range of $ 100 \,\kms \leq \Delta v \leq 300 \,\kms$. When seen face-on, the galaxy is always detected at $> 5\sigma$; in the edge-on case it remains undetected because the larger intrinsic FWHM pushes the peak flux below the detection limit. This suggests that some of the non-detections reported for high-$z$ galaxies might be due to inclination effects.
\end{itemize}

\section*{Acknowledgements}
MK acknowledges the support from the ESO-SSDF 18/24 grant and hospitality by European Southern Observatory in Munich, where part of this work has been developed.
MK, AF and SC acknowledge support from the ERC Advanced Grant INTERSTELLAR H2020/740120. LV acknowledges funding from the European Union's Horizon 2020 research and innovation program under the Marie Sklodowska-Curie Grant agreement No. 746119.
This research was supported by the Munich Institute for Astro- and Particle Physics (MIAPP) of the DFG cluster of excellence \quotes{Origin and Structure of the Universe}.
We acknowledge use of the Python programming language \citep{VanRossum1991}, Astropy \citep{astropy}, Cython \citep{behnel2010cython}, Matplotlib \citep{Hunter2007}, NumPy \citep{VanDerWalt2011}, \code{Pymses} \citep{Labadens2012}, and SciPy \citep{scipyref}.
Also, we are thankful to the anonymous referee for insightful comments and valuable suggestions.

\bibliographystyle{mnras}
\bibliography{master,codes}     

\appendix

\section{CMB effect}\label{app01}

\begin{figure}
\centering
\includegraphics[width=0.5\textwidth]{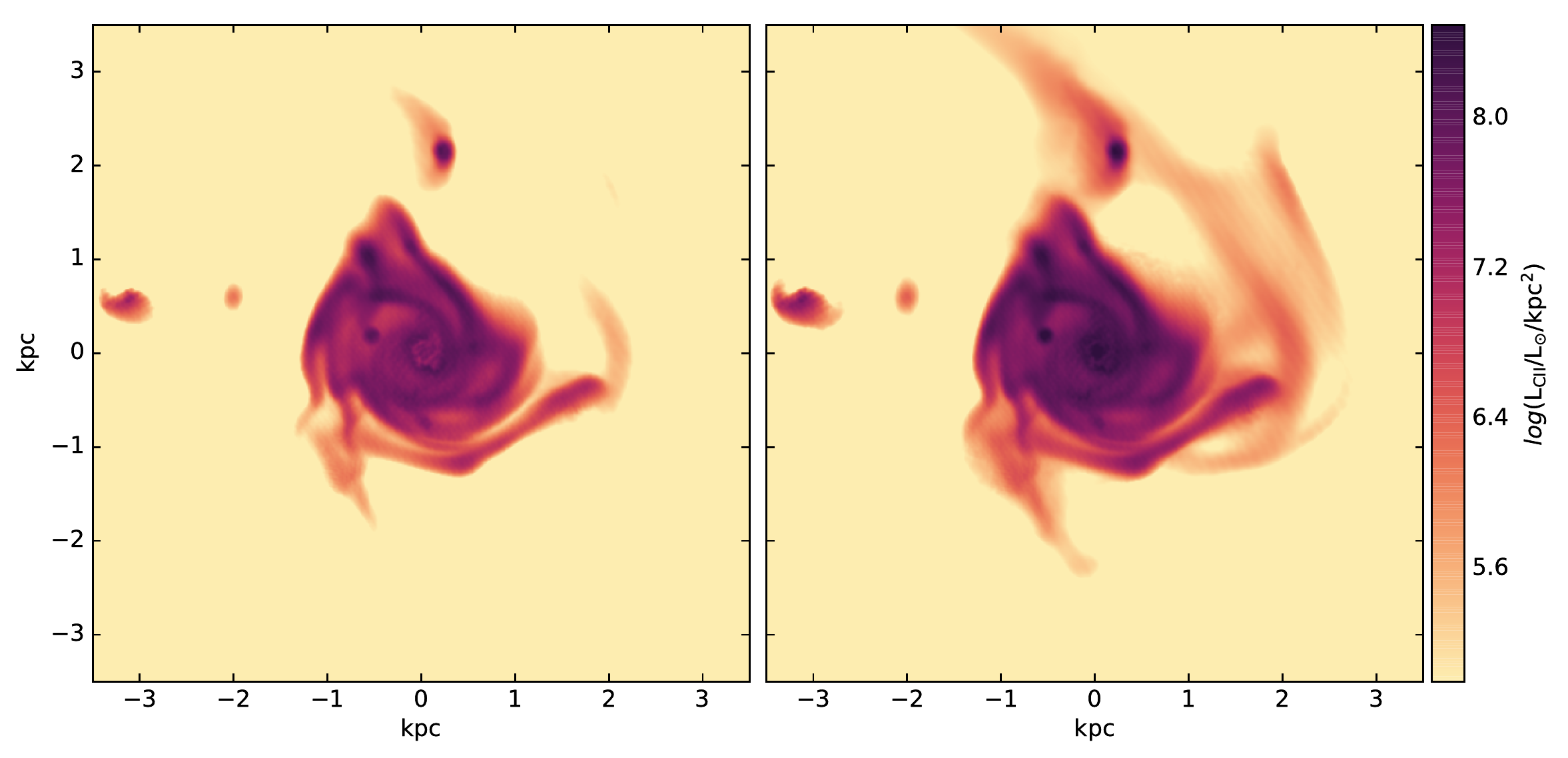}
\caption{\CII~surface brightness of \althaea~at redshift $z=6$. In the left panel ($\mathrm{L_{tot}} = 10^{8.19}\lsun $) the effect of CMB is included while in the right panel ($\mathrm{L_{tot}} = 10^{8.52}\lsun $) it is not. \label{fig05}}
\end{figure}

Suppresion due to the CMB is crucial for a correct analysis of the FIR emission coming from high redshifts \citep{DaCunha2013,Pallottini+15,vallini:2015}. In Fig. \ref{fig05} we compare the \CII~surface brightness maps with (see eq. \ref{eq:suppression_base_def}) and without ($\eta=1.0$) the inclusion of CMB suppression.
Primarily, the CMB suppresses the extended part of the signal, that is typically produced by diffuse gas ($n\lsim 5 \rm {\rm cm}^{-3}$). Note that some degree of suppression is present also for high density gas ($n\sim 100 \rm {\rm cm}^{-3}$), i.e. those dense regions that have kinetic temperature close to $\tcmb$. This fact can be understood from the trend of $\eta$ with gas density and temperature (see Fig. \ref{fig01}).
As a consequence of CMB quenching, in this specific case the total luminosity is reduced by about a factor of $\simeq 2$, i.e. from $10^{8.52}\lsun$ to $10^{8.19}\lsun$.


\bsp	
\label{lastpage}
\end{document}